\documentclass[bibyear]{aa}  

\usepackage{graphicx,color}
\usepackage{txfonts}
\usepackage[breaklinks, colorlinks, citecolor=blue]{hyperref}
\def\HII{H{\sc ii} }

\def\UC{UC~H{\sc ii}}

\def\kms{\mbox{km~s$^{-1}$}}

\def\g31{G31}
\def\stokes{Stokes\,$I$, $Q$, and $U$} 

\newcommand{\be}{\begin{equation}}
\newcommand{\ee}{\end{equation}}

\newcommand{\sigb}{$\Sigma_{B}$}
\newcommand{\mjybeam}{\,mJy\,beam$^{-1}$}

\begin{document}
\title{Self-similarity of the magnetic field at different scales: the case of G31.41+0.31} 
\author{M.\ T.\ Beltr\'an\inst{1}, M.\ Padovani\inst{1}, 
D.\ Galli\inst{1}, N.\ \'A\~nez-L\'opez\inst{2},  J.\ M.\ Girart\inst{3, 4}, R.\ Cesaroni\inst{1}, D.\ Dall'Olio\inst{5, 1}, G.\ Anglada\inst{6}, C.\ Y.\ Law\inst{7}, A.\ Lorenzani\inst{1}, L.\ Moscadelli\inst{1}, \'A.\ S\'anchez-Monge\inst{3, 4}, M.\ Osorio\inst{6}, Q.\ Zhang\inst{8}}

\institute{
INAF-Osservatorio Astrofisico di Arcetri, Largo E.\ Fermi 5,
I-50125 Firenze, Italy
\and
Universit\'e Paris-Saclay, Universit\'e Paris Cit\'e, CEA, CNRS, AIM, 91191, Gif-sur-Yvette, France
\and
Institut de Ci\`encies de l’Espai (ICE, CSIC), Can Magrans s/n, E-08193 Cerdanyola del Vall\`es, Catalonia, Spain
\and
Institut d’Estudis Espacials de de Catalunya (IEEC), E-08034 Barcelona, Catalonia, Spain
\and
Department of Space, Earth and Environment, Chalmers University of Technology, Onsala Space Observatory, Observatoriev\"agen 90, 43992 Onsala, Sweden
\and
Instituto de Astrof\'{\i}sica de Andaluc\'{\i}a, CSIC, Glorieta de la Astronomía, s/n, 18008 Granada, Spain
\and 
Department of Space, Earth \& Environment, Chalmers University of Technology, SE-412 96 Gothenburg, Sweden
\and
Center for Astrophysics | Harvard \& Smithsonian, 60 Garden Street, Cambridge, MA 02138, USA 
}
\offprints{M.\ T.\ Beltr\'an, \email{maria.beltran@inaf.it}}
\date{Received date; accepted date}

\titlerunning{Polarization in G31.41+0.31}
\authorrunning{Beltr\'an et al.}

\abstract
{Dust polarization observations of the massive protocluster G31.41+0.31 carried out at $\sim$1$''$ ($\sim3750$\,au) resolution
with the SMA at 870\,$\mu$m have revealed one of the clearest examples to date of an hourglass-shaped magnetic field morphology in the high-mass regime. Additionally, $\sim0\farcs24$ ($\sim900$\,au) resolution observations with ALMA at 1.3\,mm have confirmed these results. The next step is to investigate whether the magnetic field maintains its hourglass-shaped morphology down to circumstellar scales.}
{To study the magnetic field morphology toward the four (proto)stars A, B, C, and D contained in G31.41+0.31 and examine whether the self-similarity observed at core scales ($1''$ and $0\farcs24$ resolution) still holds at circumstellar scales, we carried out ALMA observations of the polarized dust continuum emission at 1.3\,mm and 3.1\,mm at an angular resolution of $\sim0\farcs068$ ($\sim250$\,au), sufficient to resolve the envelope emission of the embedded protostars.}
{We used ALMA to perform full polarization observations at 233\,GHz (Band 6) and 97.5\,GHz (Band 3) with a synthesized beam of $0\farcs072\times0\farcs064$. We carried out polarization observations at two different wavelengths to confirm that the polarization traces magnetically aligned dust grains and is not due to dust self-scattering.}
{The polarized emission and the direction of the magnetic field obtained at the two wavelengths are basically the same, except for an area between the embedded sources C and B. In such an area, the emission at 1.3\,mm could be optically thick and affected by dichroic extinction. In the rest of the core, the similarity of the emission at the two wavelengths suggests that the polarized emission is due to magnetically aligned grains. The polarized emission has been successfully modeled with a poloidal field with a small toroidal component on the order of 10\% of the poloidal component, with a position angle $\phi=-63^\circ$,  an inclination $i = 50^\circ$, and a mass-to-flux ratio $\lambda = 2.66$. The magnetic field axis is oriented perpendicular to the NE--SW velocity gradient detected in the core. The strength of the plane-of-the-sky component of the mean magnetic field, estimated using both the Davis-Chandrasekhar-Fermi and the polarization-intensity gradient methods, is in the range $\sim$10--$80$\,mG, for a density range $1.4 \times 10^7$--$5 \times 10^8$\,cm$^{-3}$. 
The mass-to-flux ratio is in the range $\lambda\sim1.9$--3.0, which suggests that the core is ``supercritical''. The polarization-intensity gradient method indicates that the magnetic field cannot prevent gravitational collapse inside the massive core. The collapse in the external part of the core is (slightly) sub-Alfvénic and becomes super-Alfvénic close to the center.
}
{Dust polarization measurements from large core scales to small circumstellar scales, in the hot molecular core G31.41+0.31 have confirmed the presence of a strong magnetic field with an hourglass-shaped morphology. This result suggests that the magnetic field could have a relevant role in regulating the star-forming process of massive stars at all scales, although it cannot prevent the collapse. However, it cannot be ruled out that the large opacity of the central region of the core may hinder the study of the magnetic field at  circumstellar scales. Therefore, high-angular resolution observations at longer wavelengths, tracing optically thinner emission, are needed to confirm this self-similarity.}

\maketitle

\section{Introduction}

Magnetic fields, along with gravity and turbulence are the main actors involved in the process of star formation according to theory, and depending on which of them dominates, the outcome of such a process might be different. This is especially valid for high-mass star-forming regions, where the forces involved are typically stronger than those in low-mass regions (e.g., Tan et al.~\cite{tan14}; Beltr\'an \& de Wit~\cite{beltran16}). According to Tang et al.~(\cite{tang19}), the relative importance of magnetic field, gravity, and turbulence over different physical scales drives the different modes of fragmentation seen at sub-pc scales, with a decreasing level of fragmentation expected for
increasing magnetic field strength
(e.g.,  Commer\c{c}on et al.~\cite{commercon11}, \cite{commercon22}).  However, while turbulent and gravitational energies are measured relatively easily via observations of line and dust continuum emission, characterizing the magnetic field is more challenging. 

Linearly polarized dust emission observations are the most used technique to probe the magnetic field morphology in star-forming regions (Hull \& Zhang~\cite{hull19}), since elongated dust grains are aligned with their minor axes parallel to the magnetic field as a result of radiative torques (RATs) 
(see Lazarian et al.~\cite{lazarian15}). 
An hourglass-shaped magnetic field configuration is the natural outcome of the gravitational dragging of an
initially relatively uniform field, either during the quasi-static phase of evolution of a molecular cloud core 
controlled by ambipolar diffusion (e.g. Mouschovias et al.~\cite{mouschovias06}), or during the phase of dynamical collapse 
of a cloud in the process of forming a star or a stellar cluster (Galli \& Shu~\cite{galli93a}, \cite{galli93b}).
At circumstellar scales, the magnetic field morphology might deviate from such a configuration, starting to show a clear toroidal component and a spiral morphology, once rotation (kinetic energy) dominates over magnetic forces. Polarization observations at high-angular resolution searching for these magnetic field configurations, in particular the hourglass shape, have been carried out with millimeter interferometers since the late 90s, first with the Berkeley-Illinois-Maryland Association (BIMA) millimeter array (e.g., Rao et al.~\cite{rao98}), and then with the Combined Array for Research in Millimeter-wave Astronomy (CARMA) (e.g., Hull et al.~\cite{hull14}), the Submillimeter Array (SMA) (e.g, Girart et al.~\cite{girart06}), and the Atacama Large Millimeter/submillimeter Array (ALMA).  

Evidence of hourglass-shaped magnetic fields were first reported toward the high-mass cloud  OMC-1 (Schleuning~\cite{schleuning98}) and the low-mass core NGC\,1333 IRAS\,4A (Girart et al.~\cite{girart99}). However, these observations had poor resolution ($\gtrsim3\farcs5$) and sensitivity, and did not allow to properly resolve the morphology of the magnetic field at core scales. The first textbook case of an hourglass-shaped magnetic field at core scales was observed toward the low-mass core NGC\,1333 IRAS\,4A with the SMA at 870~$\mu$m and $\sim$1$\farcs$5 by Girart et al.~(\cite{girart06}), while in the high-mass regime, the first clear example is that of the hot molecular core (HMC) G31.41+0.31 (hereafter G31) observed with the SMA at 879~$\mu$m and $\sim$1$''$ by Girart et al.~(\cite{girart09}). In both cases, the analysis of the interferometric polarization data shows that the magnetic field morphology is consistent with the prediction in the standard core collapse models for magnetized clouds (Galli \& Shu~\cite{galli93a},\cite{galli93b}; Fiedler \& Mouschovias~\cite{fiedler93}; Gon\c{c}alves et al.~\cite{goncalves08}; Frau et al.~\cite{frau11}). Subsequently, in particular with the advent of ALMA, the hourglass morphology has been reported in other low-mass cores (e.g., Davidson et al.~\cite{davidson14}; Kwon et al.~\cite{kwon18}; Maury et al.~\cite{maury18}) and in massive cores (e.g., Qiu et al.~\cite{qiu14}; Li et al.~\cite{li15}).

Observations at circumstellar scales (few 100s au) obtained at very high-angular resolution with ALMA have revealed a possible problem with the interpretation of linearly polarized dust emission. In fact,  at the typical high densities and optical depths in the disks, theory predicts that the polarized emission could also be produced by self-scattering of large dust grains (Kataoka et al.~\cite{kataoka15}), with self-scattering being the dominant polarization mechanism for grain sizes $\gtrsim$50~$\mu$m (see Kataoka et al.~\cite{kataoka17}). Dust scattering has been observed in both low- and  high-mass circumstellar disks (Girart et al.~\cite{girart18}; Bacciotti et al.~\cite{bacciotti18}; Hull et al.~\cite{hull18}; Dent et al.~\cite{dent18}), preventing the study of the magnetic field close to the protostars. However, in a few cases, like e.g. in the low-mass star-forming region BHB07--11, the polarized emission appears to trace the magnetic field at disk scales (Alves et al.~\cite{alves18}), and this made it possible to model the magnetic field with a poloidal plus a toroidal component produced by the disk rotation, as predicted by theory. Since the dust self-scattering efficiency is wavelength dependent, the best way to distinguish between different polarization mechanisms is to carry out observations at different wavelengths: if the polarization maps show no dependence on wavelength, then the observations will likely be tracing the emission of magnetically aligned grains; otherwise, the observations will likely be affected by dust scattering.

\begin{figure*}
\begin{center}
\includegraphics[angle=0,width=17cm]{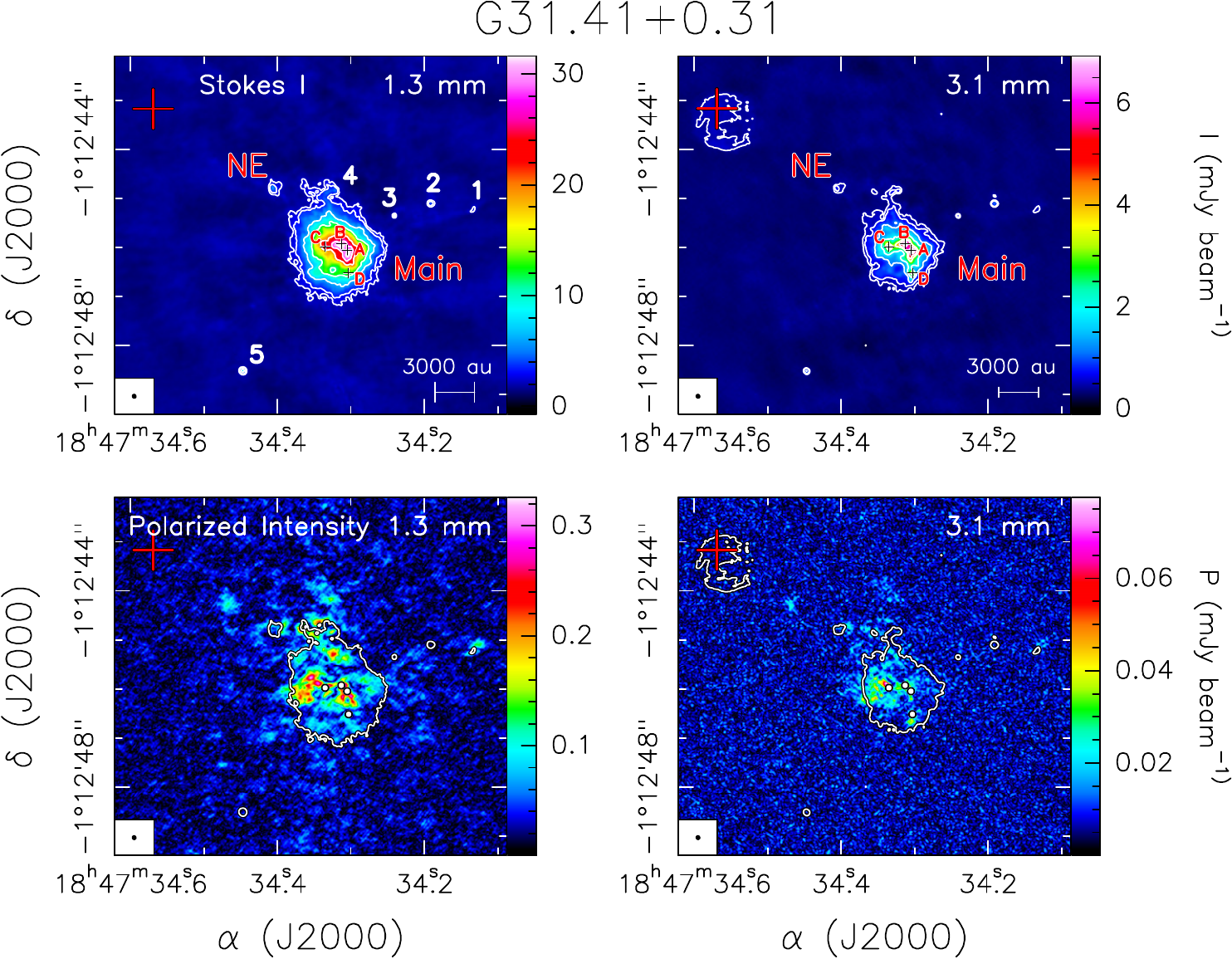}
\caption{Stokes\,$I$ and polarized intensity at both wavelengths in G31. ({\it Top panels}) ALMA Stokes\,$I$ map of the G31 HMC at 1.3\,mm ({\it left}) and 3.1\,mm ({\it right}).  The contours are $-$5, 5, 10, 30, 60, 120, 160, and 200 times $\sigma$  at 1.3\,mm and  $-$5, 5, 10, 30, 60, 120, 160, 240 and 320 times  $\sigma$ at 3.1\,mm, where 1$\sigma$ is 0.15\,mJy\,beam$^{-1}$ and 0.02\,mJy\,beam$^{-1}$, at  1.3\,mm and 3.1\,mm, respectively. Crosses ({\it top panels}) and white dots ({\it lower panels}) mark the position of four continuum sources embedded in the Main core, named A, B, C, and D, observed at 1.4\,mm and 3.5\,mm (Beltr\'an et al.~\cite{beltran21}). White numbers indicate the other continuum sources identified in the region by Beltr\'an et al.~(\cite{beltran21}). The red cross indicates the peak of the \UC\ region imaged by Cesaroni et al.~(\cite{cesa94}). The synthesized beam is shown in the lower left corner. ({\it Bottom panels})   Linearly polarized intensity $P$ ({\it colors}) and dust continuum emission map ({\it contours}) at 1.3 mm ({\it left}) and 3.1\,mm ({\it right}). 1$\sigma$ is 8.1\,$\mu$Jy\,beam$^{-1}$ and 2.6\,$\mu$Jy\,beam$^{-1}$, at  1.3\,mm and 3.1\,mm, respectively. The contours indicate the 5$\sigma$ level of Stokes\,$I$ at each wavelength.}
\label{fig-poli}
\end{center}
\end{figure*}

The HMC G31 is one of the clearest examples of an hourglass-shaped magnetic field in the high-mass regime. This object is located at 3.75\,kpc (Immer et al.~\cite{immer19}), has a luminosity of $\sim$5$\times10^4$\,$L_\sun$ (Osorio et al.~\cite{osorio09}), and displays a clear NE--SW velocity gradient suggestive of rotation (e.g., Beltr\'an et al.~\cite{beltran04}), accelerating infall (Girart et al.~\cite{girart09}; Mayen-Gijon et al.~\cite{mayen14}; Beltr\'an et al.~\cite{beltran18}; Estalella et al.~\cite{estalella19}) and rotational spin-up (Beltrán et al.~\cite{beltran18}). The dust emission of the HMC was first resolved into two massive cores, the Main and the NE core  with ALMA at $\sim$0$\farcs22$ ($\sim$825\,au) resolution (Beltr\'an et al.~\cite{beltran18}). The Main core is the most massive one, with a mass of $\sim$70\,$M_\sun$ (Cesaroni~\cite{cesa19}) and it has been recently resolved into a small protocluster composed of at least four massive sources, named A, B, C, and D, within the central 1$''$ ($\sim$3750\,au) region of the core. The sources have masses ranging from $\sim$15 to $\sim$26\,$M_\sun$ (Beltr\'an et al.~\cite{beltran21}) and are all associated with signatures of infall and outflows (Beltr\'an et al.~\cite{beltran22}).  The four dust continuum sources embedded in the Main core have also been detected at centimeter wavelengths (Cesaroni et al.~\cite{cesa10}; Beltr\'an et al.~\cite{beltran21}), and their emission is probably associated with thermal radio jets. As mentioned above, SMA 879\,$\mu$m observations at 1$''$ ($\sim$3750\,au) resolution have shown that the magnetic field lines threading the G31 HMC are pinched along its major axis, resulting in the characteristic hourglass shape (Girart et al.~\cite{girart09}). These observations have also revealed that the magnetic field dominates centrifugal and turbulent forces in the dynamics of the collapse. Additionally, Beltr\'an et al.~\cite{beltran19}) carried out ALMA 1.3\,mm polarization observations at 
a higher angular resolution of $\sim$0$\farcs$24 ($\sim$900\,au) that allowed us to better trace the magnetic field in the Main core. Those observations revealed that the magnetic field morphology was similar to that observed at 879\,$\mu$m and 1$''$ resolution. A possible explanation is that magnetic field remains important relative to gravity and turbulence from 0.02\,pc to 0.004\,pc scales and, therefore, magnetic field orientations are correlated, similar to what was observed by Zhang et al.~(\cite{zhang14}) in their SMA polarization survey of high-mass star-forming regions. 
The magnetic field morphology traced at $\sim$0$\farcs$24 has been successfully modeled by Beltr\'an et al.~(\cite{beltran19}) with a semi-analytical magnetostatic model of an envelope supported by magnetic fields with uniform mass-to-flux ratio $\lambda$=2.66 (Li \& Shu~\cite{li96}; Padovani \& Galli~\cite{padovani11}). The best-fit model suggests that the magnetic field is well represented by a purely poloidal field, with a small toroidal component of the order of 10\% of the
poloidal component, which suggests that the rotation of the core has little effect on the magnetic field (Beltr\'an et al.~\cite{beltran19}).  

The magnetic field strength in G31 estimated by  Beltr\'an et al.~(\cite{beltran19}) with the Davis-Chandrasekhar-Fermi (DCF) method (Davis~\cite{davis51}; Chandrasekhar \& Fermi~\cite{chandra53}) is $\sim$10\,mG and it is one of the highest ever measured at core scales. However, despite the magnetic field being important in G31, it is not sufficient to prevent fragmentation and collapse of the core, as demonstrated by the presence of (at least) four sources embedded in the Main core, and the fact that infall is accelerating toward the center of the core. All this suggests that gravity dominates over magnetic forces and that the self-similarity of the magnetic field observed at core scales might be produced by the large opacity of the dust emission that prevents the detection of any inhomogeneity in the core. To trace the magnetic field down to circumstellar (disk/jet) scales and study the magnetic field morphology toward the embedded (proto)stars, we carried out ALMA 1.3\,mm and 3.1\,mm polarization observations at $\sim$0$\farcs068$ ($\sim$250\,au). The goal was to investigate whether the self-similarity observed at 1$''$ (3750\,au) and 0$\farcs$24 ($\sim$900\,au) resolution still holds at disk scales. If the magnetic field dominates over turbulent and centrifugal forces, then its shape should be poloidal, as observed at large scales, whereas if the centrifugal forces dominate at the smallest scales then the magnetic field should become more toroidal. Observations at both wavelengths were carried out to investigate whether the polarization is due to the magnetic field or to dust self-scattering, which should introduce a dependence on the wavelength. 

In this work, we analyze these very high-angular resolution polarization observations, and model the magnetic field with the same model used by Beltr\'an et al.~(\cite{beltran19}). The article is organized as follows: in Sect.~2 we describe the ALMA observations; in Sect.~3 we present the properties of the polarized emission; in Sect.~4 we model the magnetic field and estimate its strength using different techniques; in Sect.~5 we discuss the properties of the magnetic field at different scales in G31, and its importance with respect to other forces involved in the star-formation process.

\section{Observations}

Interferometric full polarization observations of \g31 were carried out in Band 3 and Band 6 with ALMA in Cycle 6 as part of project 2018.1.00632.S (P.I.: M.\ Beltr\'an). Observations in Band 3 were centered at 97.5\,GHz and were carried out in the C43–8 array configuration, while those in Band 6 were centered at 233\,GHz and with the C43–7 array configuration. The total observing time in Band 3 was divided into five different execution blocks, with three of them observed on July 12, 2019,  and two of them observed on August 8, 2021. The observing time in Band 6 was divided into two execution blocks observed on July 18, 2019. We used ALMA in full polarization mode and observed all four cross correlations using a spectral setup with four $\sim$1875\,MHz spectral windows in FDM mode.  From the (XX, XY, YX, and YY) visibilities we obtained the Stokes\,$I$, $Q$, and $U$ in the image plane.  The baselines of the observations range from $\sim$47 to $\sim$12645\,m in Band 3 and from $\sim$92 to $\sim$8548\,m in Band 6.

The phase reference center of the observations is $\alpha$(J2000)$=18^{\rm h}\,47^{\rm m}\,34\fs308$, $\delta$(J2000)$=-01^\circ\,12^\prime\,45\farcs90$. Phase calibration was performed using quasar J1851+0035, while flux and bandpass calibrations were performed using quasar J1924$-$2914. Quasars J1733$-$1304 and J1751+0939 in Band 3 and quasar J1751+0939 in Band 6 were observed to determine the instrumental contribution to the cross-polarized interferometer response. 

The data were calibrated and imaged using the {\sc CASA}\footnote{The {\sc CASA} package is
available at \url{http://casa.nrao.edu/}} software package (McMullin et al.~\cite{mcmullin07}). Following ALMA Memo 5993\footnote{\url{https://library.nrao.edu/public/memos/alma/main/memo599.pdf}}, we conservatively assumed that the uncertainties on the absolute flux density calibration were $\sim$6\% at 3.1 mm and $\sim$10\% at 1.3 mm. Maps  were created using the {\tt tclean} task with the {\tt robust} parameter of Briggs~(\cite{briggs95}) of 0.5.  Because the angular resolution of the observations at 3.1\,mm and 1.3\,mm is very similar, to properly compare the emission at both wavelengths, the final maps were produced with the same {\tt uv}-coverage. This allowed us to obtain maps with the same synthesized beam, which is $0\farcs072\times0\farcs0.064$ at a position angle
PA of 50$^\circ$. The rms noise of the maps is 20\,$\mu$Jy\,beam$^{-1}$ for Stokes\,$I$ and 3\,$\mu$Jy\,beam$^{-1}$ for Stokes\,$Q$ and $U$
in Band 3 and 150\,$\mu$Jy\,beam$^{-1}$ for Stokes\,$I$ and 15\,$\mu$Jy\,beam$^{-1}$ for Stokes\,$Q$ and $U$
in Band 6. The fact that the rms noise of Stokes\,$I$ is a factor of $\sim$7 (3.1\,mm) and $\sim$10 (1.3\,mm) higher than that of Stokes\,$Q$ and $U$ is due to a problem of imaging dynamic range, because the dynamic range for Stokes\,$I$ is $>200$ at both wavelengths.  Further imaging and
analysis were done with the {\sc GILDAS}\footnote{The {\sc GILDAS} package is available at
\url{http://www.iram.fr/IRAMFR/GILDAS}} software package.  

From the \stokes, we have derived the linearly polarized intensity, $P=\sqrt{Q^2+U^2}$, the fractional linear polarization, $p=P/I$, and the polarization position angle, $\psi=\frac{1}{2}\arctan(U/Q)$.  The accuracy of the polarization position angle  $\psi$ is $\lesssim 1^\circ$ near the beam center and $\sim$1--5$^\circ$ near the full-width at half-maximum area of the primary beam (Hull et al. \cite{hull20}), while that of the  fractional linear polarization $p$ is $\sim 0.1$\%. Assuming that the polarization is produced by magnetically aligned dust grains, in all the figures we show polarization segments rotated by $90^\circ$ to outline the orientation of the magnetic field.

\begin{figure}
\begin{center}
\includegraphics[angle=0,width=\columnwidth]{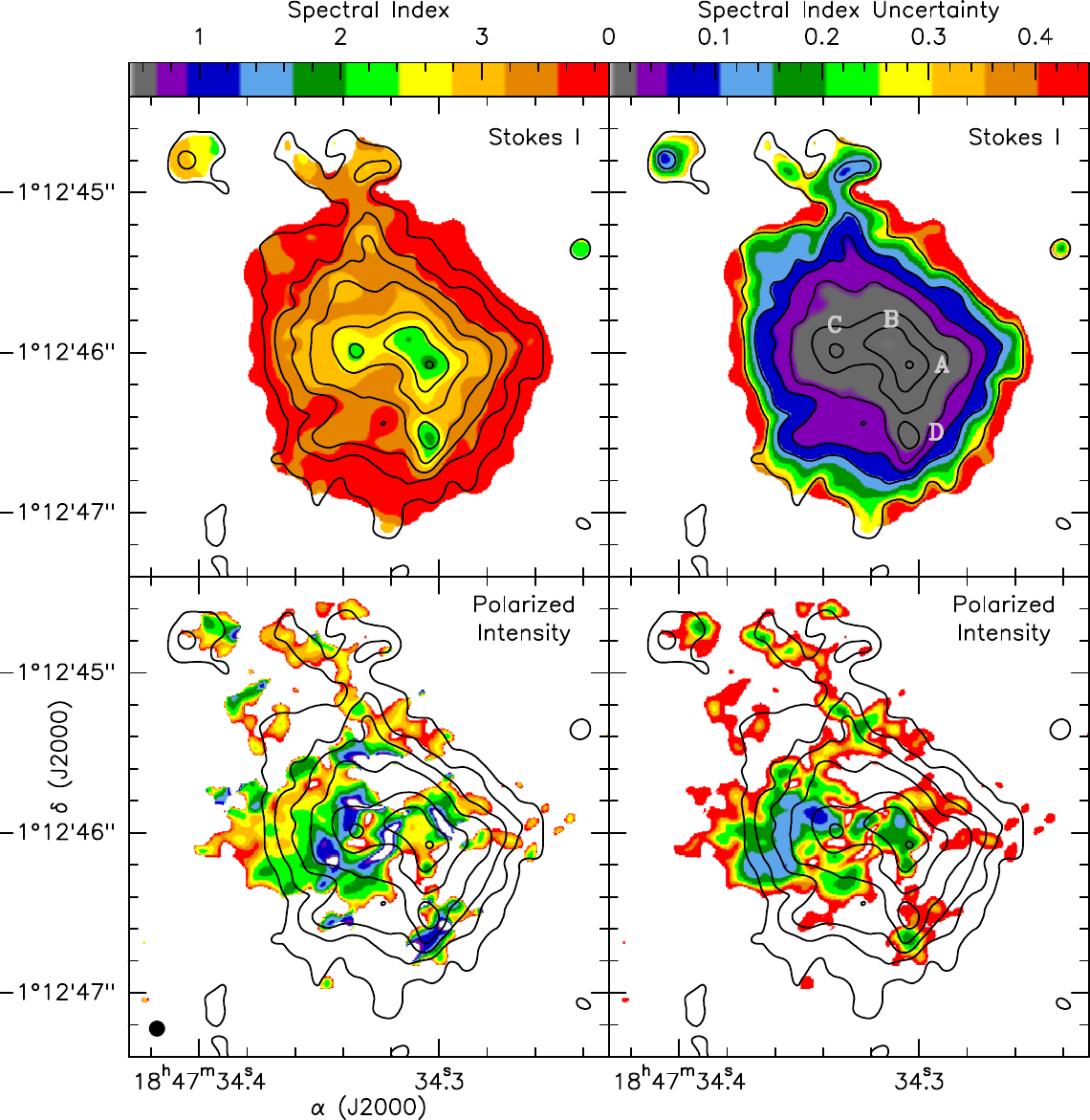}
\caption{Spectral indices. Color images of the spectral indices between 3.1 and 1.3\,mm of the Stokes\,$I$ ({\it left top panel}) and of the polarized intensity ({\it left bottom panel}). The associated uncertainties images are shown in the right panels . The scale is the same for the two images (shown in the wedge at the top of the panels). The contours in both panels show the Stokes $I$ emission at 3.1\,~mm. Contours are 1, 3, 6, 12, 24, 48 and 96\% of the peak intensity, 12~mJy~beam$^{-1}$. The images presented here were obtained using a common {\tt uv}-range for the 3.1 and 1.3~mm data (see text).}
\label{fig-specindex}
\end{center}
\end{figure}

\section{Results}

\subsection{Continuum emission}

Figure~\ref{fig-poli} shows the map of the Stokes\,$I$, namely, the total intensity, in G31 at two different wavelengths, 1.3\,mm and 3.1\,mm. These high-angular resolution observations, which trace spatial scales of $\sim$250\,au, have clearly resolved, especially at 3.1\,mm, the Main core in G31 into four embedded sources. These sources were first resolved by the ALMA dust continuum emission  observations at 1.4\,mm and 3.5\,mm of Beltr\'an et al.~(\cite{beltran21}), which were carried out with slightly worse angular resolution ($\sim0\farcs098$ at 1.4\,mm and $\sim0\farcs075$ at 3.5\,mm). These authors named the sources A, B, C, and D.  The other core in the region, named NE, is also visible at both wavelengths. As seen in Fig.~\ref{fig-poli}, the ultra-compact (UC) \HII\ region first imaged by Cesaroni et al.~(\cite{cesa94}) and located to the northeast of cores Main and NE is clearly visible at 3.1\,mm. The maps also clearly show other sources detected in the region by Beltr\'an et al.~(\cite{beltran21}), and identified with white numbers in Fig.~\ref{fig-poli}.

\begin{table*}
\caption[] {Position, flux densities, and peak brightness temperature of sources A, B, C, and D, embedded in the Main core of G31.41+0.31.}
\label{table-flux}
\begin{center}
\begin{tabular}{lcccccc}
\hline
&\multicolumn{2}{c}{Position$^a$}
\\
 \cline{2-3} 
&\multicolumn{1}{c}{$\alpha({\rm J2000})$} &
\multicolumn{1}{c}{$\delta({\rm J2000})$} &
\multicolumn{1}{c}{$I^{\rm peak}_{\rm 1.3mm}$} &
\multicolumn{1}{c}{$I^{\rm peak}_{\rm 3.1mm}$} &
\multicolumn{1}{c}{$T_{\rm B}^{\rm 1.3mm}$} &
\multicolumn{1}{c}{$T_{\rm B}^{\rm 3.1mm}$} 
\\
\multicolumn{1}{c}{Source} &
\multicolumn{1}{c}{h m s}&
\multicolumn{1}{c}{$\degr$ $\arcmin$ $\arcsec$} &
\multicolumn{1}{c}{(mJy/beam)} & 
\multicolumn{1}{c}{(mJy/beam)} & 
\multicolumn{1}{c}{(K)}  &
\multicolumn{1}{c}{(K)}  
\\
\hline
A &18 47 34.304 &$-$01 12 46.08  &30.3$\pm3.0$ &6.9$\pm0.4$ &160$\pm16$  &223$\pm13$\\
B &18 47 34.313 &$-$01 12 45.94  &31.6$\pm3.2$ &6.0$\pm0.4$ &167$\pm17$ &194$\pm13$ \\
C &18 47 34.335 &$-$01 12 46.00  &23.4$\pm2.3$ &3.7$\pm0.2$ &124$\pm12$ &120$\pm6$\\
D &18 47 34.304 &$-$01 12 46.53  &11.7$\pm1.2$ &2.6$\pm0.2$  &62$\pm6$ &84$\pm6$\\
\hline
\end{tabular}
\end{center}
\tiny
$^a$ Position of the dust emission peak at 3.1\,mm.  \\
\end{table*}

Table~\ref{table-flux} lists the position, peak flux density, $I^{\rm peak}_\lambda$, and peak brightness temperature, $T_{\rm B}$, of sources A, B, C, and D at both wavelengths. The peak brightness temperature has been estimated from the flux density at the peak of the emission following the expression $T_{\rm B} = 1.2221\times10^6/(\theta^2\,\nu^2)\,I^{\rm peak}_\lambda$, where $\theta$ is full width at half power of the synthesized beam in arcsec, $\nu$ is the rest frequency in GHz, and $I^{\rm peak}_\lambda$ is in Jy/beam. The position of the sources has been measured at the peak of the emission at 3.1\,mm, because the sources are better resolved at this wavelength, and  coincides with that measured  by Beltr\'an et al.~(\cite{beltran21}). The peak intensities (in mJy/beam) at 1.3\,mm are slightly lower than those estimated at 1.4\,mm by Beltr\'an et al.~(\cite{beltran21}), but this is consistent with the fact that the synthesized beam at 1.3\,mm is smaller. In fact, the brightness temperatures are slightly higher at 1.3\,mm. On the other hand, the intensities at 3.1\,mm and 3.5\,mm are quite similar, especially for sources A and B, as one would expect taking into account that the synthesized beams of both observations were almost the same.   As seen in  Table~\ref{table-flux}, sources A and B are the brightest at both wavelengths, and have similar peak intensities.

\begin{figure*}
\begin{center}
\includegraphics[angle=0,width=17cm]{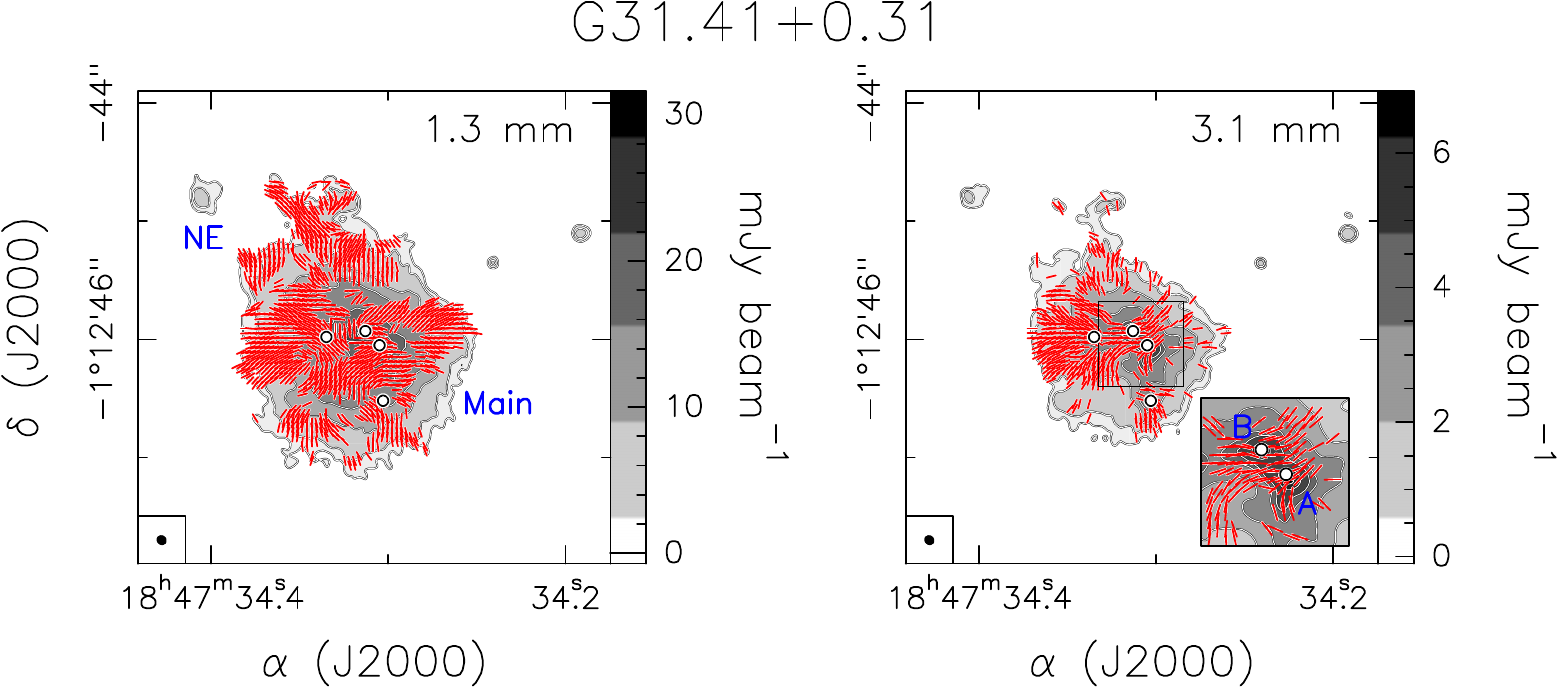}
\caption{Magnetic field orientation at both wavelengths. Magnetic field segments ({\it red lines}) at 1.3\,mm ({\it left}) and 3.1\,mm ({\it right}) overlaid on the corresponding Stokes\,$I$ maps ({\it contours} and {\it greyscale}). Segments are shown every five pixels. The white dots mark the position of four embedded continuum sources observed at 1.4\,mm and 3.5\,mm (Beltr\'an et al.~\cite{beltran21}). The black rectangle ({\it right})) indicates the zoomed region shown in the inset. Contours are the same as in Fig.~\ref{fig-poli}. The synthesized beam is shown in the lower left corner.}
\label{fig-pola}
\end{center}
\end{figure*}

\begin{figure}
\begin{center}
\includegraphics[angle=-90,width=10cm]{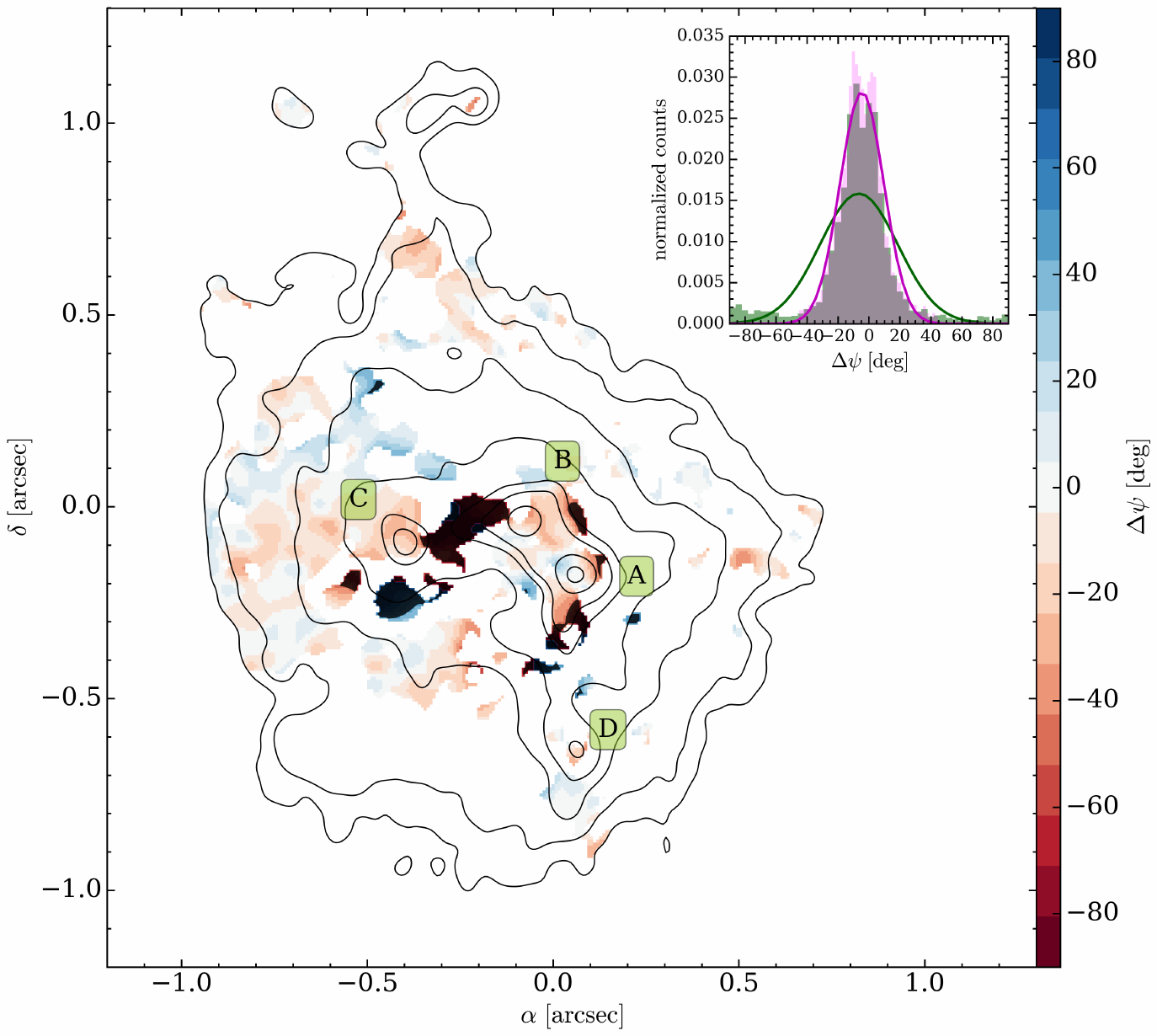} 
\caption{Difference between the polarization angles at 1.3\,mm and 3.1\,mm. Polarization angle residuals ({\it colors}) between the 1.3\,mm and 3.1\,mm observations,
superposed to the 3.1 mm dust continuum emission ({\it contours}).  Contour levels are the same as in Fig.~\ref{fig-poli}.
The letters (A,B,C,D) show the position of four embedded continuum sources listed in Table~\ref{table-flux}.
The hatched areas show the regions where $|\Delta\psi|>45^\circ$.
The inset shows the distribution of the polarization angle residuals obtained considering the whole range of $\Delta\psi$ ({\em green})
and limited to $|\Delta\psi|\leq45^\circ$ ({\em magenta}). The solid lines show the corresponding Gaussian fits to the histograms 
whose mean value and standard deviation are $-6.43^\circ\pm25.22^\circ$ and $-4.45^\circ\pm14.15^\circ$, respectively.}
\label{fig-histogram}
\end{center}
\end{figure}

\begin{figure*}
\centering
\includegraphics[angle=0,width=17cm]{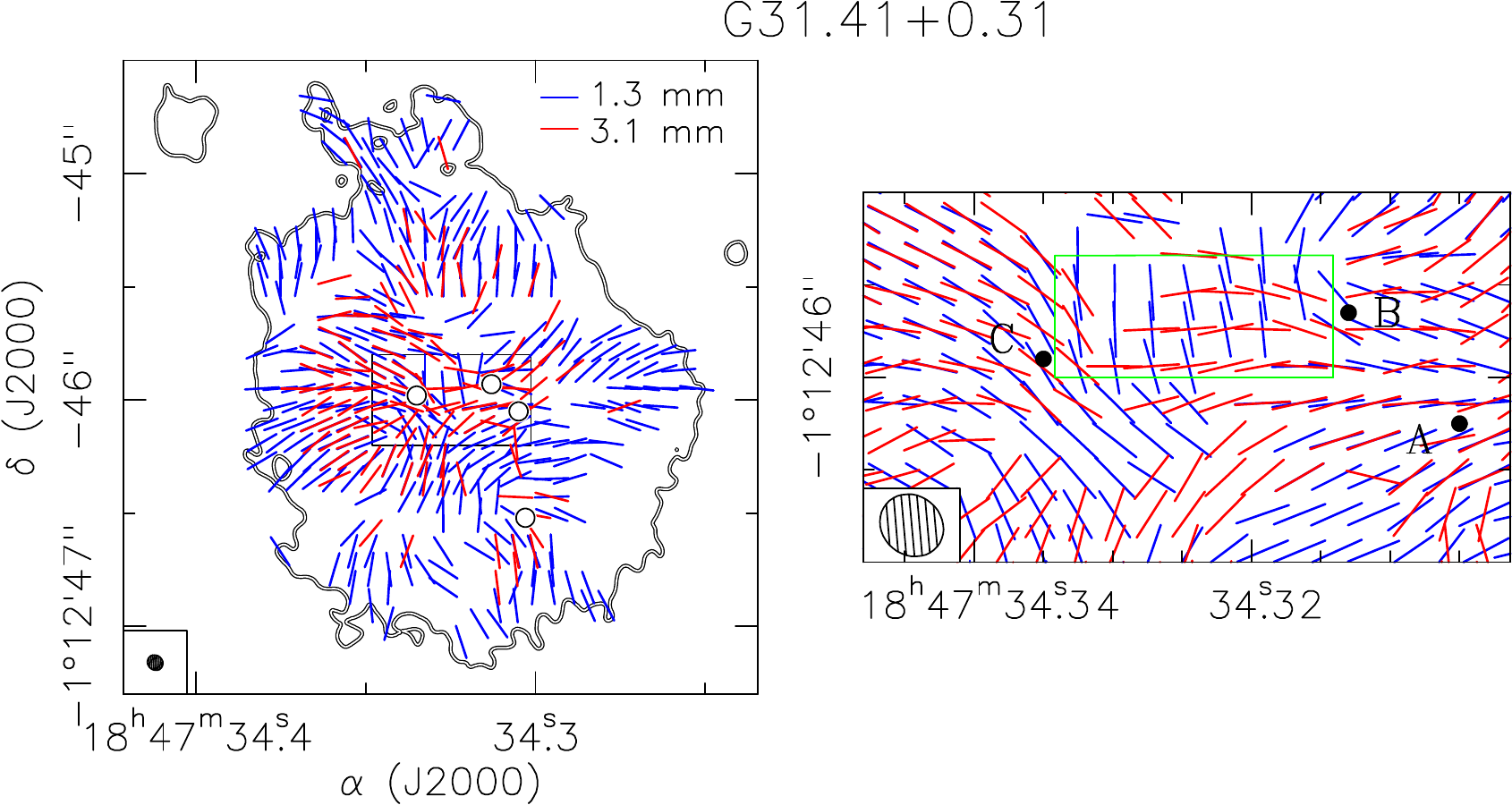}
\caption{Comparison of magnetic field orientations at 1.3\,mm and 3.1\,mm. ({\it Left}) Magnetic field segments at 1.3\,mm ({\it blue}) and 3.1\,mm ({\it red}). Segments shown every ten pixels. Contours show the 5$\sigma$-level dust continuum emission at 1.3\,mm. The synthesized beam is shown in the lower left corner. The black rectangle indicates the zoomed region shown in the right panel. ({\it Right}) Close-up of the central region toward sources A, B, and C. Segments are shown every five pixels. The green rectangle shows the region where the polarization segments at 1.3\,mm and 3.1\,mm are almost perpendicular.}
\label{fig-angle-comparison}
\end{figure*}

\begin{figure}
\begin{center}
\includegraphics[angle=0,width=8.5cm]{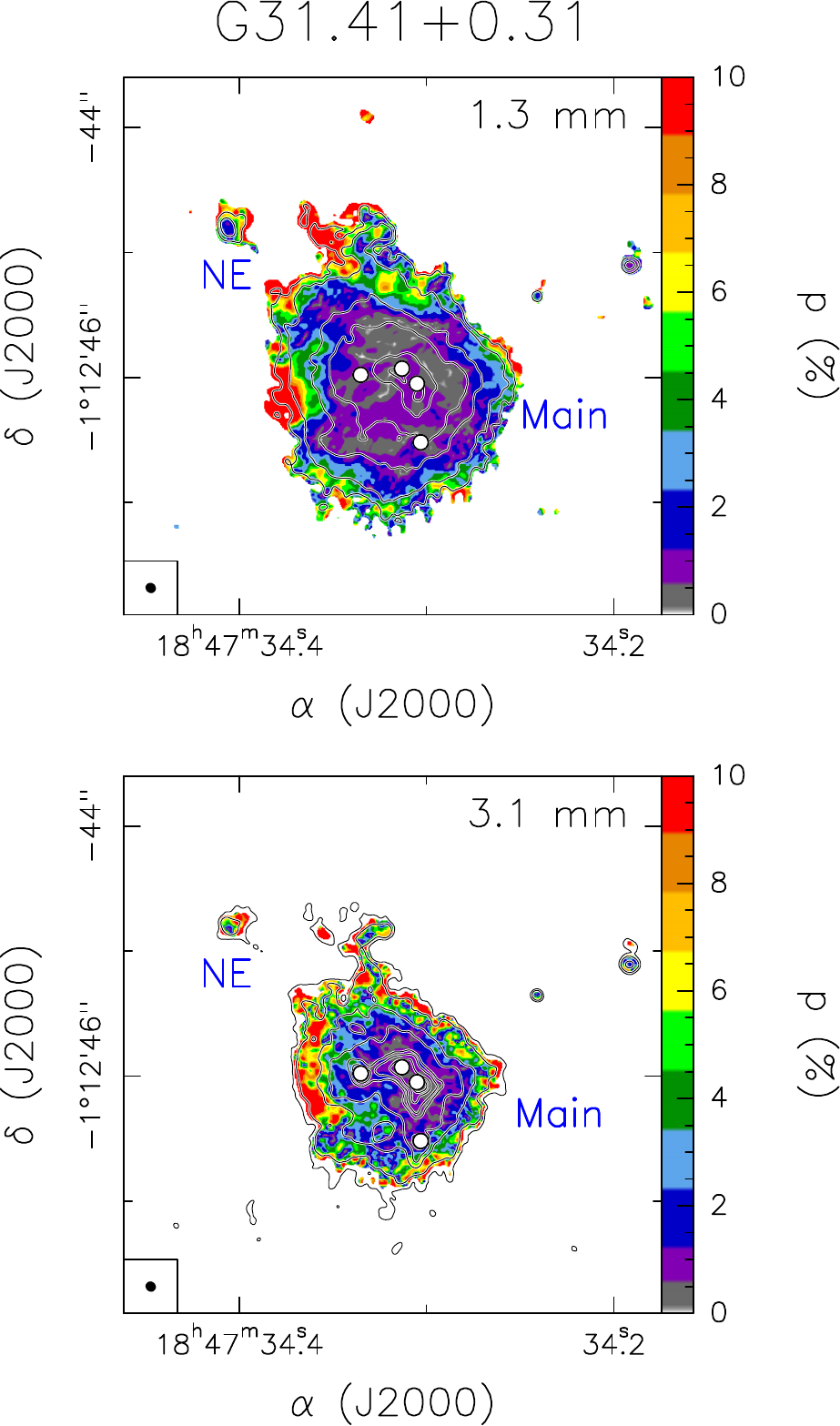}
\caption{Polarization fraction at both wavelengths. ({\it Top}) Polarization fraction $p$ ({\it colors}) and dust continuum emission map ({\it contours}) at 1.3\,mm. The synthesized beam is shown in the lower left
corner. Contours and symbols are the same as in Fig.~\ref{fig-poli}. 
({\it Bottom}) Same as top panel, at 3.1\,mm. 
}
\label{fig-polm}
\end{center}
\end{figure}

\subsection{Polarized emission}
\label{poli}

Figure~\ref{fig-poli} shows the linearly polarized emission, $P$, at both wavelengths. The polarized emission at 1.3\,mm is quite similar to the one obtained at lower angular resolution ($\sim0\farcs2$) by Beltr\'an et al.~(\cite{beltran19}). These authors show that the main peak of $P$ is eastward of source C. At our higher angular resolution, the peak of $P$ is also found to the east of source C, but these new observations have clearly resolved the emission as compared to the $P$ map of Beltr\'an et al.~(\cite{beltran19}). Secondary peaks to the north of source C, and toward the position of sources A and B are also visible at both $\sim0\farcs2$ and $0\farcs07$ angular resolution. The $P$ map at 1.3\,mm also shows some polarized detections associated with source 1, which again is also visible in the lower resolution map of Beltr\'an et al.~(\cite{beltran19}). 

The polarized emission at 3.1\,mm is weaker than at 1.3\,mm and is only visible
to the east of source C, associated with sources A, B, and D and very weakly with source 1. As seen in Fig.~\ref{fig-poli}, $P$ at 3.1\,mm is not detected to the north and to the west of the Main core, where, instead, polarized emission is seen at 1.3\,mm. On the contrary, the polarized emission associated with source D is much weaker at 1.3\,mm than at 3.1\,mm.  

\subsection{Spectral index map}

Figure~\ref{fig-specindex} presents a map of the spectral index $\alpha$ and its uncertainty, where $S_\nu\propto \nu^\alpha$, computed using the 3.1 and
1.3\,mm Stokes\,$I$ ({\it top panels}) and the polarized emission ({\it bottom panels}).  As seen in this figure, the spectral map is completely different if calculated from Stokes\,$I$ or from $P$. 
For Stokes\,$I$, there is a gradual decrease of the spectral index toward the center, from $\simeq 4.0$ on the outer part of the envelope down to 2.0 at the peak of the four massive protostars (source B has a spectral index of $1.87\pm0.02$). Given the uncertainties (from 0.2--0.3 at the border down to $\ll0.1$ at the center), this spectral index gradient is significant. The temperature in the core at the observed scales is expected to be $\sim 100$~K or more (Osorio et al. \cite{osorio09}; Beltr\'an et al.~\cite{beltran18}). Therefore, the dust continuum emission is in the Rayleigh-Jeans regime. This means, that in the regions where the emission is optically thin, $\alpha$ is related to the power-law index $\beta$ of the dust opacity coefficient $\kappa_\nu\propto \nu^\beta$, through $\alpha$ = $\beta$ + 2. At the peak position of sources A, B, C and D, the spectral index indicates optically thick emission. This suggests that the spectral index increase at larger radius could be due to a smooth decrease of the optical depth. In this case, Fig.~\ref{fig-specindex} suggests a $\beta$ of 1.0--1.5, which is the typical value in molecular clouds (e.g., D'Alessio et al.~\cite{dalessio01}; Sadavoy et al.~\cite{sadavoy13}). Alternatively, the observed variation of the spectral index across the core could be due to an increase of $\beta$ at outer radius, or by a combination of both.  Because the four embedded sources are optically thick,  we can obtain a lower limit of the dust temperature $T_d$ from the peak brightness temperature $T_B$ of  the sources (see Table~\ref{table-flux}).  $T_B$ is high for all sources embedded in the Main core, on the order of $\sim$100--200\,K. If the peak emission arises from a compact, only partially resolved emission, then $T_d$ should  be higher, which is consistent with typical temperatures of embedded massive young stellar objects such as sources A, B, C, and D (Beltr\'an et al.~\cite{beltran21}). 

In the regions of the Main core where it was possible to calculate the $P$ spectral index, the values obtained are different, clearly lower than those obtained from Stokes\,$I$. Similarly to what observed for Stokes\,$I$, the spectral index appears to decrease toward the inner part of the core, but it is not as clear as in Stokes\,$I$. This may be due to the lower signal-to-noise ratio of the emission.  Outside the center, the values range between 2 and 3 (the typical uncertainty is in most cases between 0.1 and 0.3). Near the center, the spectral index appears to be between 1.2 and 2.5. This is not expected if the polarized signal follows the same opacity law as the dust emission (see Sect.~\ref{dichroic}).

\subsection{Polarization angles}
\label{pol-angles}

Figure~\ref{fig-pola} shows the magnetic field orientation at 1.3\,mm and 3.1\,mm obtained by rotating by 90$^\circ$ the polarization segments. From now on, when we mention magnetic field orientation, we refer to this (although see Sect.~\ref{dichroic}).  The polarization angles have been calculated for Stokes\,$Q$ and  Stokes\,$U>3\sigma$. As seen in this figure, the magnetic field is better sampled at 1.3\,mm. This is consistent with the fact that the observations at 1.3\,mm trace better the polarized emission in the core, as seen in Fig.~\ref{fig-poli} and already mentioned in Sect.~\ref{poli}. The orientations of the magnetic field lines at 1.3\,mm coincide with those at 3.1\,mm over the whole core (Fig.~\ref{fig-pola}) except for an area between sources C and B that we discuss below.  To better distinguish regions exhibiting significant dispersion, in Fig.~\ref{fig-histogram} we show the map of the polarization angle residuals, 
obtained from the difference between the polarization angles at 1.3\,mm and 3.1\,mm,
as well as the corresponding histogram. As seen in this figure, in a few areas, especially between sources C and B, the polarization angle residuals are $|\Delta\psi|>45^\circ$.  The histogram of the polarization angle residuals shows that the average difference between the orientation of the magnetic field at the two wavelengths is $-6.43^\circ\pm25.22^\circ$ considering the whole range of $\Delta\psi$, and $-4.45^\circ\pm14.15^\circ$ if limited to $|\Delta\psi|\leq 45^\circ$. The fact that the polarization maps and the magnetic field orientations are basically the same at the two  wavelengths indicates that the polarized observations are probably tracing the emission of magnetically aligned grains and are not affected by dust self-scattering.

The orientation of the magnetic field lines coincides with that probed at 1.3\,mm and $\sim0\farcs24$ angular resolution by Beltr\'an et al.~(\cite{beltran19}) and is also consistent with the orientation observed at 870\,$\mu$m with the SMA at an angular resolution of 1$''$ (Girart et al.~\cite{girart09}), which trace a region $\sim$15 times larger than that traced by the ALMA 0$\farcs$068 observations. This agreement on the global orientation of the magnetic field points to a self-similarity of the magnetic field from large to small scales. 

The high-angular resolution achieved with these ALMA observations has allowed us to better trace the magnetic field lines toward the position of the embedded compact sources. As seen in Fig.~\ref{fig-poli}, the strongest Stokes\,$I$ emission is located toward sources A and B. The highest emission level contours at both 1.3\,mm and 3.1\,mm delineate a flattened structure surrounding both sources.  The magnetic field as better probed at 3.1\,mm is almost perpendicular to this flattened central region of the core (see inset in Fig.~\ref{fig-pola}), and has a morphology similar to the one observed toward the binary protostellar system  NGC\,1333\,IRAS\,4A at 870~$\mu$m by Girart et al.~(\cite{girart06}), namely an hourglass shape. The spatial scale traced by the observations in NGC\,1333\,IRAS\,4A is of $\sim$350\,au, which is quite similar to the one traced by our observations ($\sim$250\,au).   

In Fig.~\ref{fig-angle-comparison}, we have simultaneously plotted the polarization segments showing the magnetic field  orientation at 1.3\,mm and 3.1\,mm. As seen in the right panel, which zooms in the inner part of the Main core between sources B and C, the magnetic field orientation at 1.3\,mm is almost perpendicular to that at 3.1\,mm. This can also be seen in the histogram of Fig.~\ref{fig-histo-dichroic}, showing the average difference between the orientation of the magnetic field at 1.3\,mm and 3.1\,mm only for the area between sources B and C. As seen in this plot, for most of the segments $\Delta\psi$ is $>75^\circ$ or $<-75^\circ$. As discussed in Sect.~\ref{dichroic}, one plausible explanation for this behavior could be that the polarized emission at 1.3\,mm in that area is affected by dichroic extinction, or alternatively, that the polarized emission at 1.3\,mm and 3.1\,mm is probing different depths within the core.

\subsection{Polarized fraction}

Figure~\ref{fig-polm} shows the polarized fraction $p$ at both wavelengths. The polarized fraction  is $<2$\% toward the central part of the Main core and of the NE core.  As clearly seen at 3.1\,mm, this low polarized fraction coincides with the region surrounding the dust continuum embedded sources. In the rest of the core, $p$ increases a little bit and reaches values of $\gtrsim$10\%. 
As already noted by Beltr\'an et al.~(\cite{beltran19}), there is no significant increase in the polarized fraction toward the peaks of polarized intensity, especially to the east of the core.

\section{Analysis}

\subsection{Modeling the magnetic field}
\label{model}

From an initial analysis of the polarization maps at 1.3 mm and 3.1 mm, we noted a trend similar to that found from previous observations at 1.3\,mm by Beltr\'an et al.~(\cite{beltran19}). For this reason, we decided to use the same model used by these authors. The model is based on Li \& Shu~(\cite{li96}) and Padovani \& Galli~(\cite{padovani11}) and describes an axially symmetric singular toroid threaded by a poloidal magnetic field. In order to take into account the impact of rotation, we incorporated a modified force-free toroidal magnetic field component as described in Padovani et al.~(\cite{padovani13}).
While we refer to Sect.~4.1 of Beltr\'an et al.~(\cite{beltran19}) for the details of the model, here we recall the main parameters such as the mass-to-flux ratio ($\lambda$), the ratio between the toroidal and poloidal component of the magnetic field ($b_0$), the orientation of the projection of the magnetic axis on the plane of the sky measured from north to east ($\varphi$), and the inclination of the magnetic field with respect to the plane of the sky ($i$), which is assumed to be positive (negative) if the magnetic field in the northern sector points toward (away from) the observer.

To begin with, we fixed the values of $\lambda$ and $b_0$ at $\lambda=2.66$ and $b_0=0.1$, as done in Beltr\'an et al.~(\cite{beltran19}),
and used {\tt DustPol} (Padovani et al.~\cite{padovani12}) to create images of the Stokes parameters $I$, $Q$ and $U$ for each combination of $i$ and $\varphi$. Models were then fed into the {\tt simobserve} and {\tt simanalyze} tasks of the CASA software, using the same antenna configuration as in our observations. Finally, the convolved models have been compared to the two new sets of observations by computing the reduced chi squared, $\bar\chi^2$.

\begin{figure}
\begin{center}
\vspace{-1cm}
\includegraphics[angle=0,width=8.5cm]{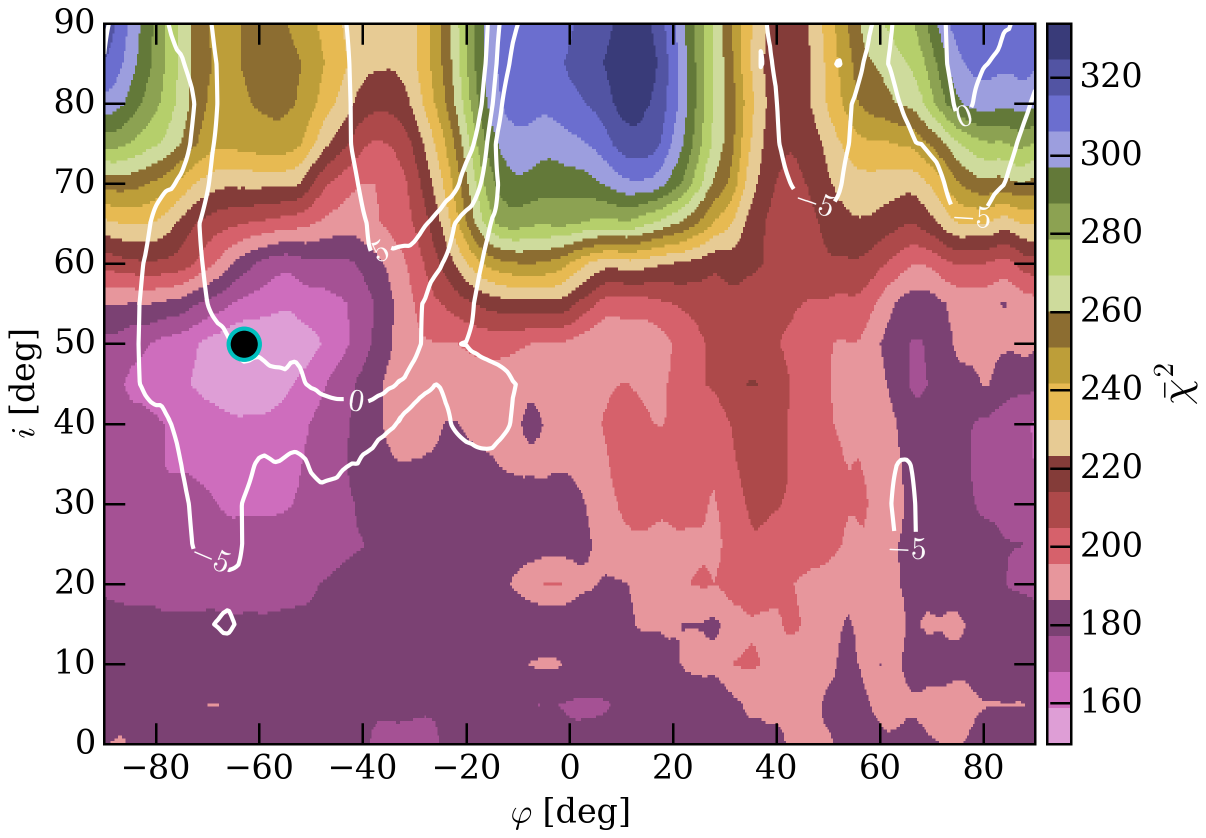}
\vspace{-3cm}
\caption{Reduced chi-squared values, $\bar\chi^2$, for 1.3 mm observations as a function of the inclination ($i$) and the orientation of the projection of the magnetic axis on the plane of the sky ($\varphi$). Black isocontours show values of the difference between the observed and modeled position angles, $\Delta\psi$, between $-10^\circ$ and $10^\circ$ in steps of $5^\circ$. The black-filled cyan circle shows the position of the $\bar\chi^2$ minimum.}
\label{fig-iphichi2-1mm}
\end{center}
\end{figure}

\begin{figure}
\begin{center}
\vspace{-3cm}
\includegraphics[angle=0,width=8.5cm]{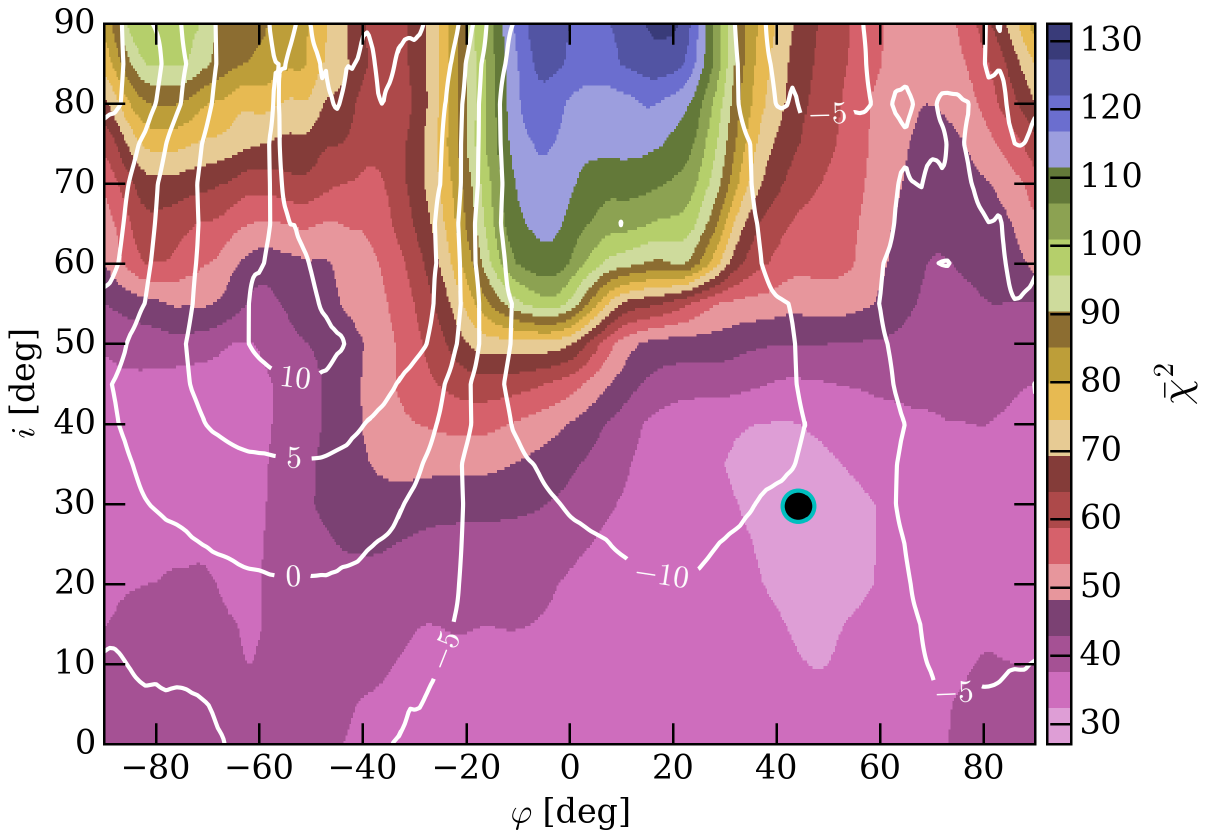}
\vspace{-3cm}
\caption{Same as Fig.~\ref{fig-iphichi2-1mm}, but for 3.1 mm observations.pdf}
\label{fig-iphichi2-3mm}
\end{center}
\end{figure}

Figure~\ref{fig-iphichi2-1mm} shows the distribution of $\bar\chi^2$ at 1.3 mm as a function of $i$ and $\varphi$.  Since the model is symmetric with respect to the mid-plane, that is with respect to $i=0^\circ$, we show results only for $i\in[0,90]^\circ$. The minimum $\bar\chi^2$ is achieved for $\varphi=-63^\circ$, close to what we found with previous observations ($\varphi=-44^\circ$). As for the inclination, here we obtain $i=50^\circ$ against the previous value of $-45^\circ$. However, the symmetry of the model does not allow to discriminate between positive and negative inclinations for optically thin emission, so the new results in absolute value are in agreement with the previous ones.
The situation is less clear at 3.1 mm, and as shown in Fig.~\ref{fig-iphichi2-3mm}, the distribution of $\bar\chi^2$ is much flatter. 
The minimum $\bar\chi^2$ is quite shallow and is obtained for a positive value of $\varphi$, in contrast to the expected orientation of G31. 
However, this discrepancy can easily be explained because the region is not uniformly sampled at 3.1\,mm. 
As shown in the right panel of Fig.~\ref{fig-pola}, the observations are not sensitive to the polarization of the northern and western regions of the core and the magnetic field morphology cannot be properly modeled.  To possibly increase the signal-to-noise of the 3.1\,mm observations and check whether we could recover some low-intensity extended emission, we applied a {\tt uv}-taper of $0\farcs15\times0\farcs15$ to the visibility data when running {\tt tclean}. The resulting synthesized beam of the maps is $0\farcs19\times0\farcs16$. After performing the {\tt uv}-tapering,  the distribution of $\bar\chi^2$ remains flat over the same range of inclinations, although the minimum value of $\bar\chi^2$ is now obtained for a negative value of $\varphi$ (see Fig.~\ref{fig-iphichi2-3mmuvtaper}),  in agreement with the observations at 1.3 mm and the predictions of Beltr\'an et al.~(\cite{beltran19}). For all these reasons, we decided to model only the 1.3\,mm observations. The best fit model has been obtained with $b_0=0.1$, $\lambda=2.66$, $i=50^\circ$, and $\phi=-63^\circ$ (see Fig.~\ref{fig-modobs}).

\subsection{Magnetic field strength}
\label{subsec:fieldstrength}

Following Beltr\'an et al.~(\cite{beltran19}), we used the results of our model fit to the 1.3\,mm observations to measure the magnetic field strength using the Davis-Chandrasekhar-Fermi (DCF) method (Davis~\cite{davis51}; Chandrasekhar \& Fermi~\cite{chandra53}). For this purpose, we used the dispersion of the polarization angle with respect to the model (see Sect.~\ref{model}). We considered the region inside the $5\sigma$ contour level to focus only on the Main core.  The average uncertainty on the observed polarization angles is $\delta\psi_{\rm obs}=0.5~\sigma_{QU}/\sqrt{Q^2+U^2}$, where $\sigma_{QU}=15\,\mu$Jy\,beam$^{-1}$ is the noise on the observed Stokes\,$Q$ and $U$. This expression is valid for high ($>5$) signal-to-noise ratios (e.g., Vaillancourt~\cite{vaillancourt06}), which is the case of our data. As seen in Fig.~\ref{fig-poli} ({\it bottom right panel}), the linearly polarized intensity, $P$, of the G31 core is $\gtrsim5\sigma_{QU}=0.07$\,mJy/beam$^{-1}$ in the region inside the $5\sigma$ contour of the 1.3~mm dust continuum map, and in such a region  $\delta{\rm \psi_{\rm obs}}\lesssim 4^\circ$, 
therefore, we used a histogram bin of $6^\circ$.

\begin{figure}
\begin{center}
\hspace*{-1cm} \includegraphics[angle=-90,width=11cm]{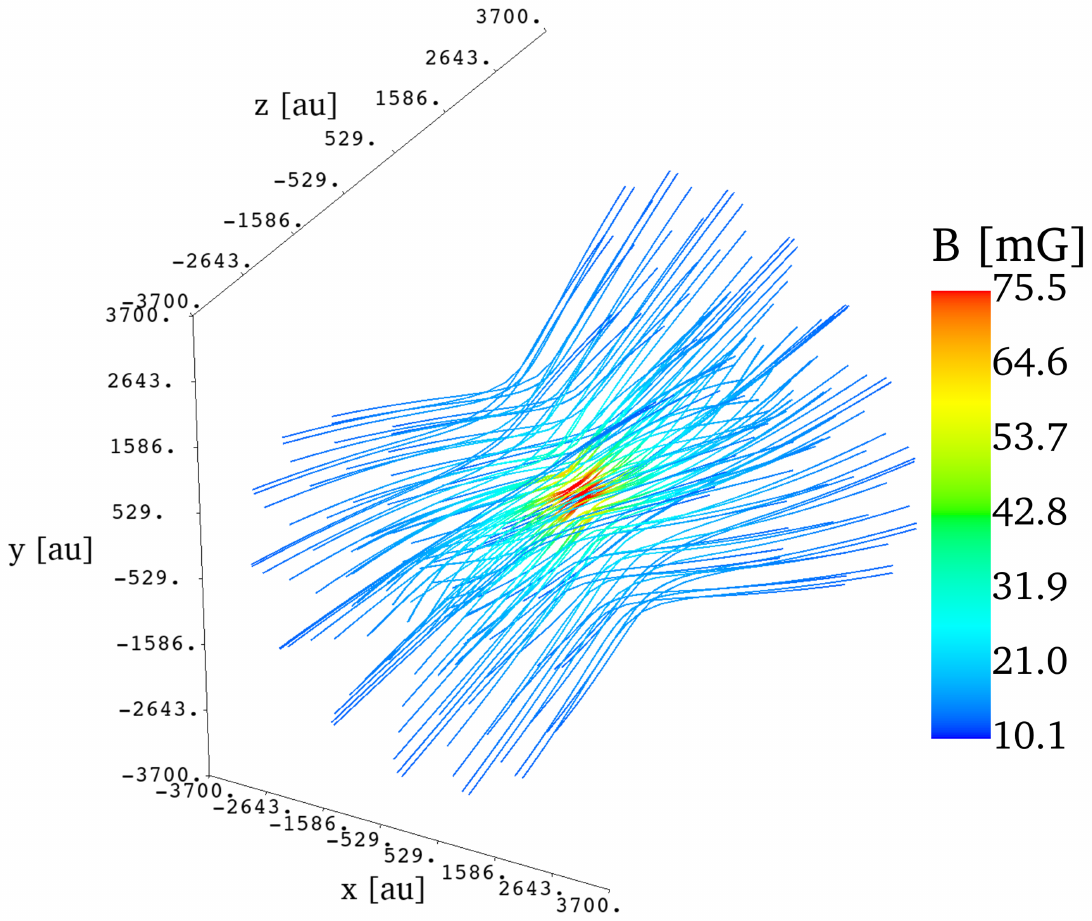}

 \vspace*{-1cm}
 \includegraphics[angle=-90,width=8.5cm]{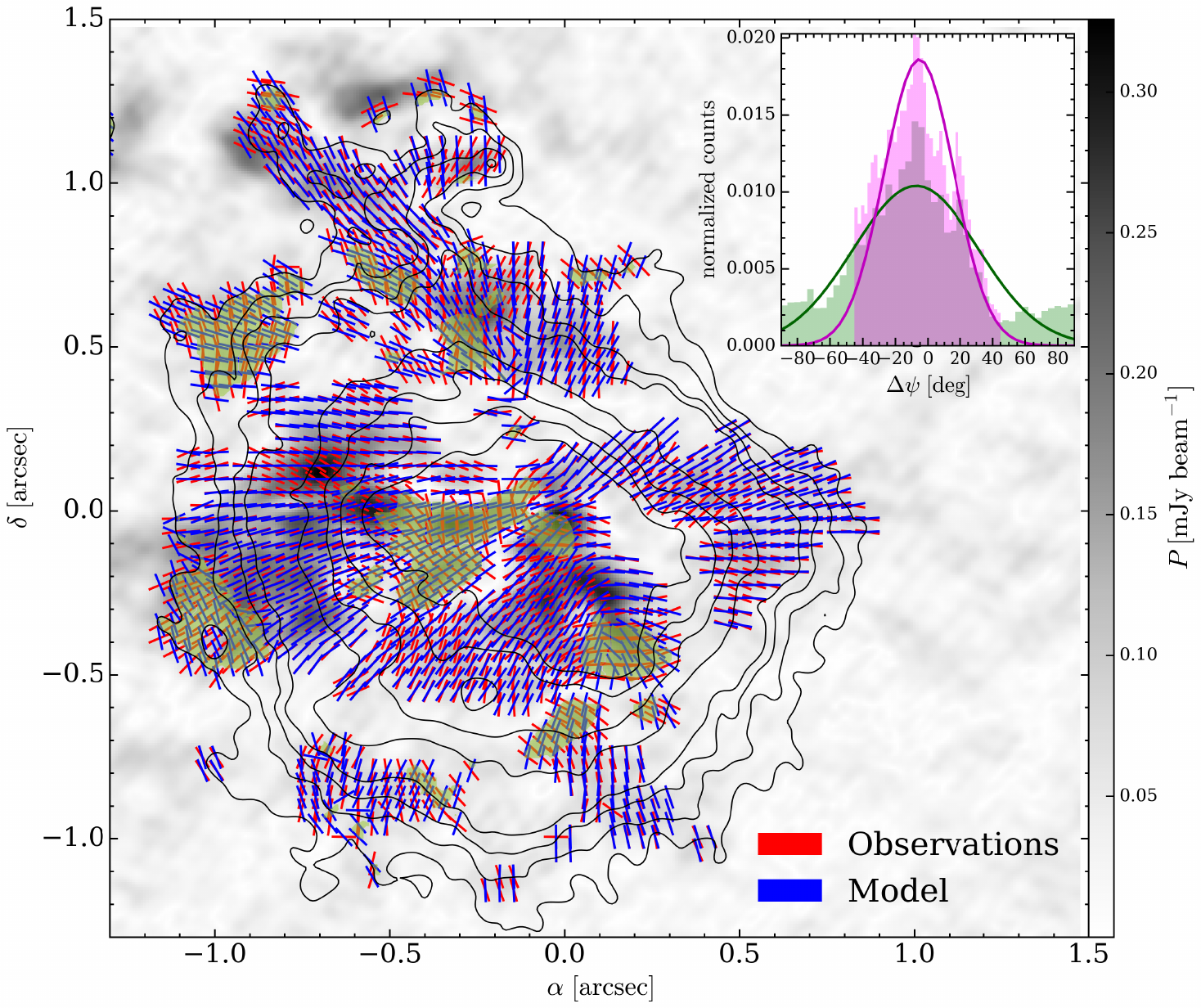}
\caption{Comparison of magnetic field orientations between model and observations. {\it Upper panel}: magnetic field configuration inside a radius of 3700\,au, corresponding to the radius of the Main core in G31, obtained with $b_0=0.1$, $\lambda=2.66$, $i=50^\circ$, and $\phi=-63^\circ$. {\it Lower panel}: Magnetic field segments from observations at 1.3\,mm ({\it red}) and the best model ({\it blue}). The grey-scale map shows the polarized intensity, $P$,
while black contours show the 1.3~mm dust emission at 5, 10, 30, 60, 120, 160, and 200 times $\sigma$, which is 0.15\,mJy\,beam$^{-1}$.
Yellow areas show the regions where the difference between the observed and modeled 
polarization angle is $|\Delta\psi|>45^\circ$.
The inset shows the distribution of the polarization angle residuals obtained considering the whole range of $\Delta\psi$ ({\em green})
and limited to $|\Delta\psi|\leq45$~deg ({\em magenta}). The solid lines show the corresponding Gaussian fits to the histograms.}
\label{fig-modobs}
\end{center}
\end{figure}

The average value of the distribution of residuals is $\Delta\psi=-7.18^\circ\pm38.41^\circ$ or $\Delta\psi=-5.32^\circ\pm21.45^\circ$ considering the whole range of $\Delta\psi$ or only that where $|\Delta\psi| < 45^\circ$,
respectively. In the area between sources B and C, the 1.3\,mm polarization observations are likely affected by dichroic extinction (see Sect.~\ref{dichroic}). If so, the 1.3\,mm data in that area do not properly trace the magnetic field and, in such an area, the polarization segments do not have to be rotated by 90$^\circ$ to recover the magnetic field orientation. This explains why the difference in angle between the observations and the model of the magnetic field obtained by us is so large, as can be seen in Fig.~\ref{fig-modobs}. For this reason, we decided to limit the range of $\Delta\psi$ between $-45^\circ$ and $+45^\circ$, and use the standard deviation on the polarization angle dispersion, $\sigma_\psi= 21.45^\circ$. 
Since the measurement uncertainty of the polarization angle $\delta\psi_{\rm obs}$ is $\lesssim 4^\circ$, 
the intrinsic dispersion is $\delta\psi_{\rm int}$ = $(\sigma_\psi^2 -  \delta\psi_{\rm obs}^2)^{1/2}\sim \sigma_\psi$.

The DCF method is based on the 
assumption that the perturbations responsible for the polarization
angle dispersion $\delta\psi_{\rm int}$ are Alfv\'en waves of amplitude $\delta B=\sqrt{4\pi\rho}\,\sigma$, where $\rho$ is the density and $\sigma$ is the velocity dispersion. This gives
\be
\label{b-strength}
B_{\rm pos}=\xi\frac{\sigma_{\rm los}}{\delta\psi_{\rm int}}\sqrt{4\pi\rho}\,,
\ee
where $B_{\rm pos}$ is the component of the magnetic field on the plane of the sky, $\sigma_{\rm los}$ is the component of the velocity dispersion along the line of sight, and $\xi=0.5$ is a correction factor derived from turbulent cloud simulations (Ostriker et al.~\cite{ostriker01}). According to Ostriker et al.~(\cite{ostriker01}), the DCF method is a valid approximation as long as $\delta\psi_{\rm int}< 25^\circ$, that would correspond to cases for which the uniform component of the magnetic field is much larger than the random components. In our case, $\delta\psi_{\rm int}=21.45^\circ$ is large but still within the limit. Note that recent simulations of Liu et al.~(\cite{liu-etal21}) indicate that, statistically, the average ratio
between the directly measured angular dispersion in polarization maps and the turbulent-to-ordered magnetic field strength ratio (i.e., the correction factor $\xi$) is $\sim$0.25 at clump and core scales if $\delta\psi_{\rm obs} < 25^\circ$, which is smaller than the value reported by Ostriker et al.~(\cite{ostriker01}).

Following Beltr\'an et al.~(\cite{beltran19}), we used ALMA observations of continuum and line emission carried out with similar angular resolutions,  $\sim0\farcs1$, to estimate $\sigma_{\rm los}$ and $\rho$. The line-of-sight dispersion has been computed from the full width half maximum $\Delta V$ of different $K$ transitions of CH$_3$CN and CH$_3^{13}$CN observed by Beltr\'an et al.~(\cite{beltran22}) as $\sigma_{\rm los}=\Delta V/\sqrt{8\,\ln{2}}$. To avoid the effects of rotation on the line broadening, we estimated $\Delta V$ at different pixel positions of the
core and then averaged the values. As already noticed by Beltr\'an et al.~(\cite{beltran19}), the thermal contribution to the velocity dispersion is negligible for the temperatures $\ga 100$\,K estimated for the Main core (e.g., Beltr\'an et al.~\cite{beltran18}). Using $\Delta V\simeq5.6\pm0.15$\,\kms, the value of $\sigma_{\rm los}$ is $2.4\pm0.06$\,\kms. This value is similar to the value of 2.1\,\kms\ estimated by Beltr\'an et al.~(\cite{beltran19}) from observations of CH$_3$CN carried out with ALMA at an angular resolution of $\sim0\farcs22$ by Beltr\'an et al.~(\cite{beltran18}). As for the mean density, Beltr\'an et al.~(\cite{beltran19}) used a value of $n=1.4\times 10^7$~cm$^{-3}$, corresponding to the volume density of the Main core averaged inside a radius of  $\sim1''$. This radius corresponds to that of the Main core inside the $5\sigma$ contour level (see Fig.~\ref{fig-poli}). Beltr\'an et al.~(\cite{beltran21}), with ALMA observations at an angular resolution of  $\sim0\farcs1$, estimated average values of the number density toward the embedded continuum sources A to D of $n\gtrsim 5\times10^9$\,cm$^{-3}$ (see their Table\,4). Because these are average densities inside the embedded sources, these values have to be taken as upper limits of the densities in the inner region of the core ($R\sim0\farcs3$) containing sources A, B, and C, which we assume to be an order of magnitude lower, $n\sim 5\times10^8$\,cm$^{-3}$. Using the range of number densities of 1.4$\times10^7$--5$\times10^8$\,cm$^{-3}$, $\sigma_{\rm los}$=2.4\,\kms, and $\delta\psi_{\rm int}=21.45^\circ$, we obtain a magnetic field strength in the plane-of-sky of $B_{\rm pos}=8.3$--50\,mG, which taking into account the inclination with respect to the plane of the sky, $i=50^\circ$, obtained from our modeling (see Sect.~\ref{model}), corresponds to a  total magnetic field strength $B=B_{\rm pos}/\cos i\sim$13--78\,mG. Note that these values are consistent with those reported in Fig.~12 of Crutcher~(\cite{crutcher12}) for a sample of molecular clouds if extrapolated to the densities of $\sim10^7$--$10^8$\,cm$^{-3}$ estimated in G31, and with those reported in Fig.~3 of Liu et al.~(\cite{liu22}).

\subsection{The polarization-intensity gradient method}
\label{subsec:polint}

As seen in the previous section, the magnetic field strength is usually estimated from maps of polarized emission using the DCF method. However, this method leads to a single value of the field strength statistically averaged  over an entire region, and does not allow to trace a position-dependent field strength. To derive a position-dependent strength based on the observed magnetic field morphology, one can use instead the polarization-intensity gradient method developed and successfully applied to observations by Koch et al.~(\cite{koch12a}, \cite{koch12b}, \cite{koch13}, \cite{koch14}). Based on the MHD force equation, this method uses the measurable angle between the direction of the field and the intensity gradient, together with the direction of gravity, to derive a local magnetic field strength at every location where polarization is detected.
The relative importance of magnetic and gravitational forces is represented by the non-dimensional quantity $\Sigma_B$, defined by
\begin{equation}
    \label{eq:sigma}
    \Sigma_{B} = \frac{\sin \Psi}{\sin \alpha}
\end{equation}
where $\Psi$ is the angle in the plane of the sky between the direction of the local gravity and the intensity gradient, and $\alpha$ is the angle  between the polarization direction and the intensity gradient. The local gravity direction is computed through the distribution of all surrounding mass, at any map position, under the assumption that the dust emission is proportional to the gas mass.
Values of $\Sigma_B$ greater or less than unity indicate whether or not the magnetic force is able to prevent gravitational collapse.

\begin{figure}
    \centering
    \includegraphics[trim= -1.5cm 1cm -1.5cm 1cm clip, width=8.5cm, angle=-90]{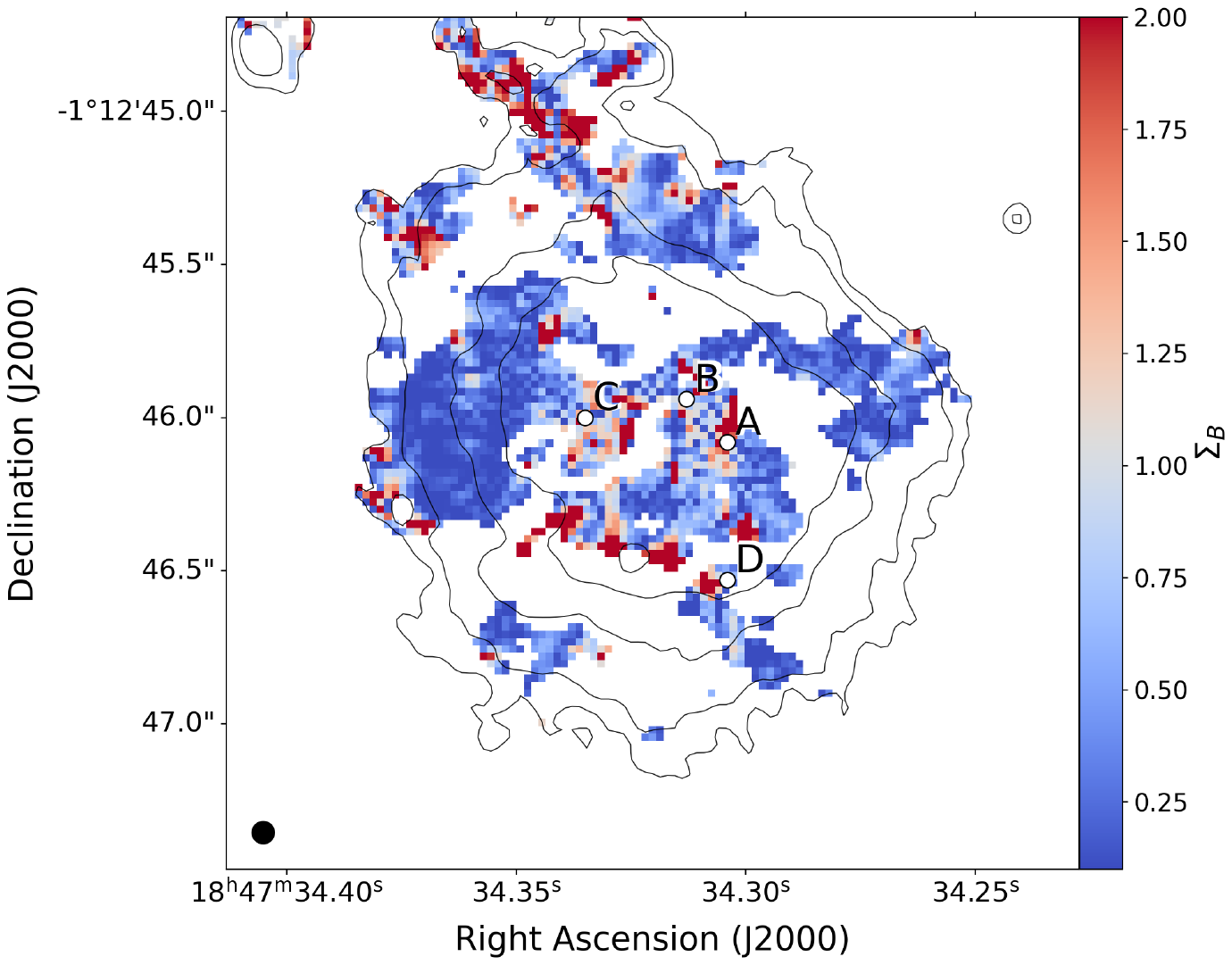}
    \caption{Map of $\Sigma_B$ at 1.3~mm ({\it color}\/) superimposed to intensity contours. Contour levels are  5, 10, 30, 60, 120, 160, 200 times $\sigma$ ($\sigma$= 0.15\mjybeam)}
    \label{fig:sigbmap}
\end{figure}

The map of $\Sigma_B$ at 1.3~mm is shown in Fig.~\ref{fig:sigbmap}.  For this calculation we have reduced the number of pixels contained in each beam surface to $<$3, and we have considered only polarization measurements above $5\sigma$.

\begin{figure}
    \centering
\includegraphics[trim= -1.5cm 1cm -1.5cm 1cm clip, width=8.5cm, angle=-90]{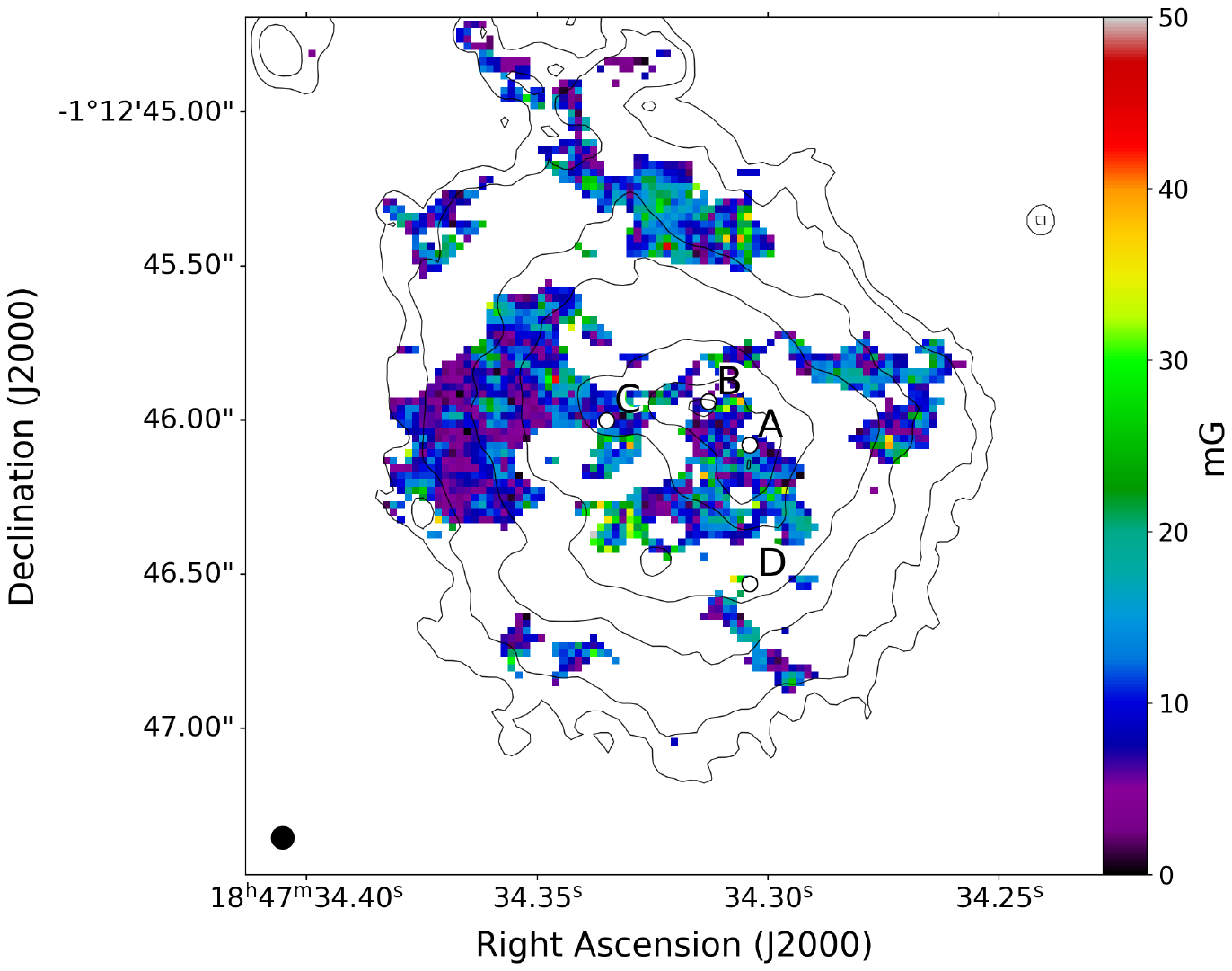}
    \caption{Magnetic field strength estimated with the polarization-intensity gradient method. Map of $B_{\rm pos}$ derived from Eq.~(\ref{eq:bfieldKoch}) for a uniform density $n=1.4\times 10^7$~cm$^{-3}$ ({\it color map}\/) superimposed to intensity contours. Contours levels are  5, 10, 30, 60, 120, 160, 200 times $\sigma$ ($\sigma$= 0.15\mjybeam)}
    \label{fig:bfieldstrength}
\end{figure}

The magnetic field strength in the plane of the sky 
is given in terms of $\Sigma_B$ by
\begin{equation}
   B_{\rm pos} =
   \sqrt{\Sigma_B \left(\nabla P + \rho \nabla \phi \right)  4\pi R_{\mathrm{c}}} ,
   \label{eq:bfieldKoch}
\end{equation}
(Koch et al.~\cite{koch12a, koch12b}), where $\nabla P$ is the pressure gradient,
$\nabla \phi$ is the gravity acceleration, 
and $R_{\mathrm{c}}$ is the radius of curvature of the magnetic field line. 
The latter is defined by
\begin{equation}
    \label{eq:curvature}
    \dfrac{1}{R_{\mathrm{c}}} = \dfrac{2}{d} \cos \left[\dfrac{1}{2} (\pi - \Delta PA) \right],
\end{equation}
where $d$ is the distance between two contiguous magnetic field segments, and $\Delta PA$ is the difference between their position angles. The radius of curvature is calculated by averaging the local curvatures between two adjacent segments over all nearest neighbors. 
In Eq.~(\ref{eq:bfieldKoch}) we assume that local changes in temperature and density are minor compared to gravity and therefore the pressure gradient $\nabla P$ to be negligible compared to the gravitational force $\rho\nabla\phi$. Using the average value of \sigb, $\nabla \phi$ and $R_{\rm c}$ over the map for Eq.~(\ref{eq:bfieldKoch}) and (\ref{eq:curvature}), 
and a uniform value of $\rho$ corresponding to 
the density range of the Main core, $n=1.4\times 10^7$--$5\times 10^8$\,cm$^{-3}$, we obtain 
$B_{\rm pos} = 9$--$55$\,mG.
This result is consistent with the range obtained with the DCF method in Sect.~\ref{subsec:fieldstrength} ($B_{\rm pos} = 8.3$--$50$\,mG).
We note that only values of $\Sigma_B < 2$ have been considered to avoid artificially high values of the magnetic field strength due to $\sin\alpha\sim 0$ values in Eq.~(\ref{eq:sigma}).  If we increase the threshold of $\Sigma_B$ to 5, the average magnetic field strength $B_{\rm pos}$ increases by $\sim$6\%. Figure~\ref{fig:bfieldstrength} presents the magnetic field strength map at 1.3~mm derived for a uniform density of $1.4\times10^7$\,cm$^{-3}$. 
The bulk of the $B_{\rm pos}$ values are in the range $\sim$0.2--50~mG.

\section{Discussion}

\subsection{Wavelength polarization dependence}

The polarization spectral index has a behaviour quite distinctive from the total dust emission.  At first, the polarization spectral index follows the same trend as the dust emission spectral emission, i.e., the value increases outwards (from the dust intensity peaks). However, the absolute values are  significantly smaller for the polarization spectral index, reaching a maximum between 2.5 and 3.0, whereas the dust spectral index maximum is between 3.0 and 3.5. The decrease with radius, could be due to the decrease of the optical depth of total dust (Hildebrand et al.~\cite{hildebrand00}; Yang et al.~\cite{yang17}), but in this case one would expect similar values where the emission is optically thin. 
Therefore, this suggests an increase of the polarization fraction, or of the polarization efficiency, with wavelength. This has also been observed in the circumbinary disk around the low mass YSO binary BHB2007-11 (Alves et al. \cite{alves18}) and in the HL\,Tau disk Lin et al. ~(\cite{lin23}).  However, the origin of this increase is not  understood. 
Further theoretical studies and more observations  are need to better understand and characterize the increase of polarized fraction with increasing wavelengths.

\begin{figure}
\begin{center}
\includegraphics[angle=0,width=10cm]{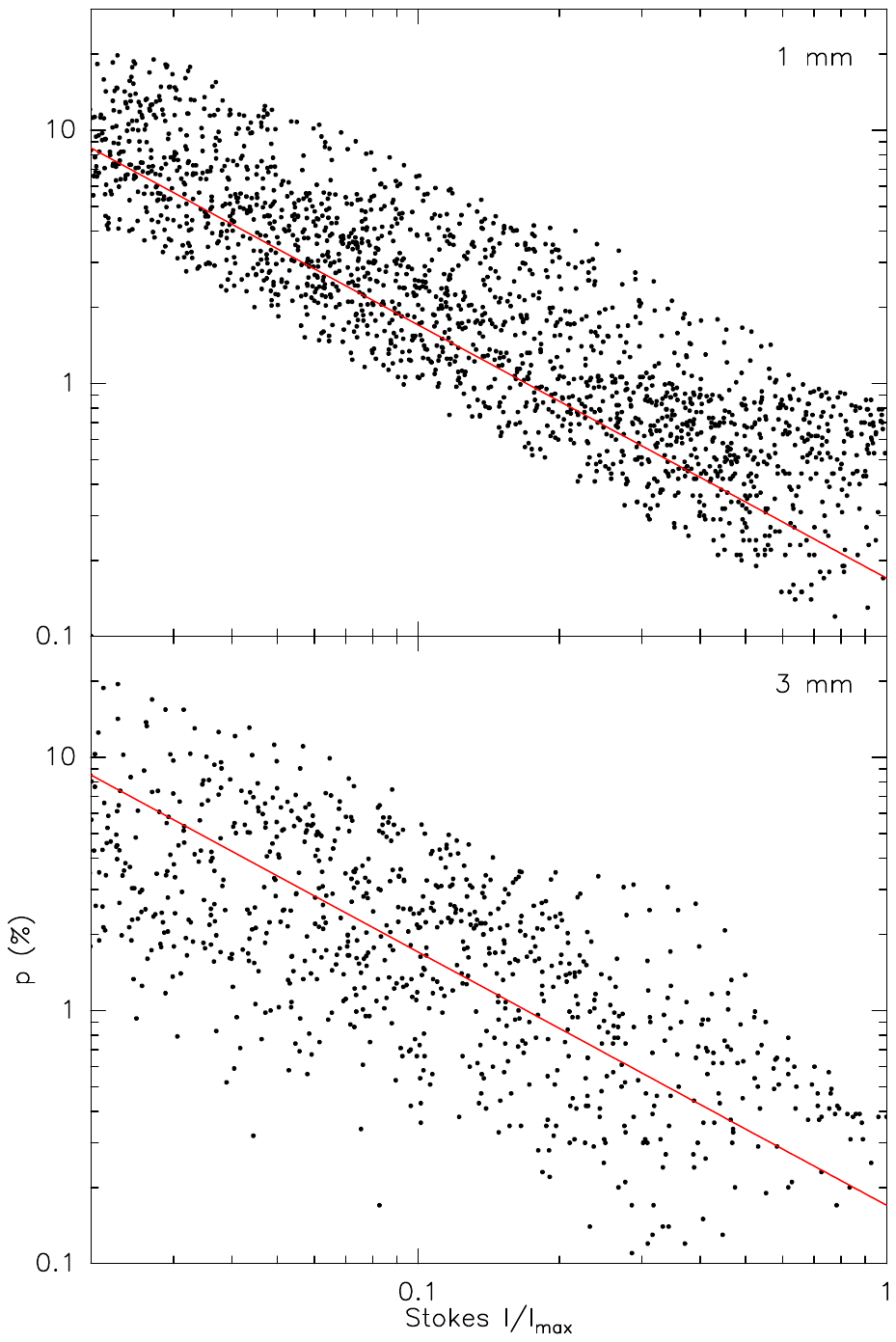}
\caption{Polarization fraction $p$ as a function of the normalized Stokes\,$I$, that is $I/I_{\rm max}$, where $I_{\rm max}$ is the maximum value of the intensity at each wavelength. The dots show the values obtained with a sampling of $0\farcs15$ that fulfill the following requirements: $I> 0.2$\,mJy\,beam$^{-1}$ and $>0.8$\,mJy\,beam$^{-1}$ at 3.1\,mm and 1.3\,mm, respectively, and polarized intensity higher than 0.03\,mJy\,beam$^{-1}$  and 0.06\,mJy\,beam$^{-1}$, respectively. The red line is only for illustration  and indicates $p=$(Stokes\,$ I/I_{\rm max})^{-1}$.}
\label{fig-I-polm}
\end{center}
\end{figure}

\subsection{Dichroic extinction}
\label{dichroic}
 
Figure~\ref{fig-angle-comparison} shows the magnetic field segments at 1.3\,mm and 3.1\,mm.  As seen in the zoom-in, the magnetic field orientations obtained from the two wavelengths differ from each other by almost 90$^\circ$ in the area located between sources B and C. This can also be seen in the histogram of Fig.~\ref{fig-histo-dichroic} which shows the average difference between the orientations of the magnetic field at 1.3\,mm and 3.1\,mm only for the area between sources B and C. One sees that for most of the segments $\Delta\psi$ is $>75^\circ$ or $<-75^\circ$. The fact that the polarization directions at the two wavelengths are almost perpendicular, suggests that in this inner area of the Main core, the polarization mechanism could be affected by  dichroic extinction, in which linear polarization originates when light travels through an optically thick medium with dust grains aligned by the magnetic field in the presence of a temperature gradient (e.g., Wood~\cite{wood97}). Dichroic extinction, which has been observed in low-mass star-forming regions (e.g., Ko et al.~\cite{ko20}; Liu~\cite{liu21}),  occurs because the component of the electric vector parallel to the long axis of the dust grains is more efficiently absorbed, and this results in polarization being parallel to the magnetic field orientation. In the case of G31, the polarized emission at 1.3\,mm between sources B and C could be optically thick and dominated by dichroic extinction, while the polarized emission at 3.1\,mm would be optically thinner and dominated by dust grains aligned by the magnetic field.  It is worth mentioning that, despite G31 being a high-mass core with high dust opacity, the area affected by dichroic extinction is very confined to a small area between sources B and C, and this is likely due to the high optical depth needed for dichroic extinction to work. 

\begin{figure*}
\begin{center}
 \includegraphics[angle=0,width=17cm]{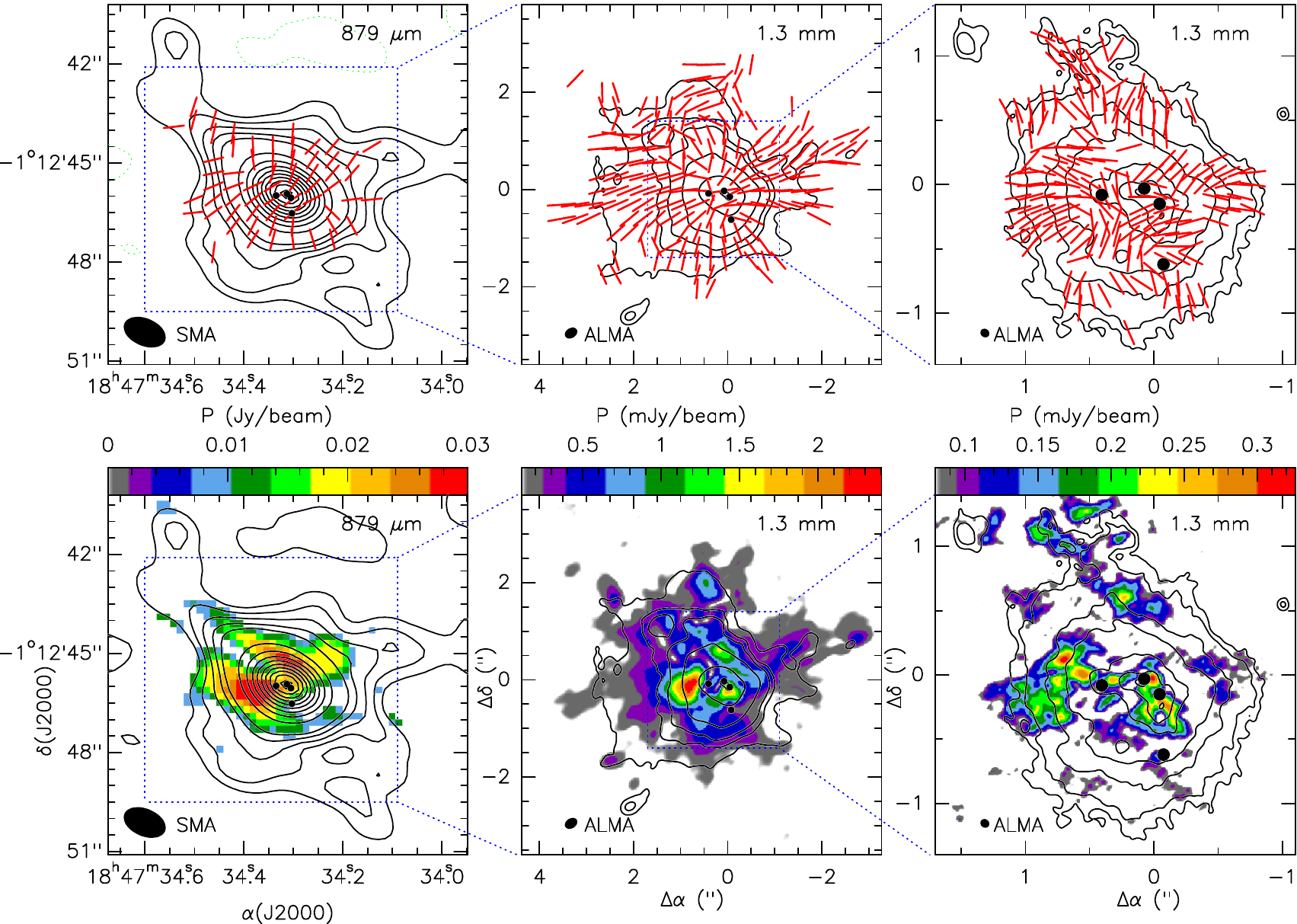}
\caption{Comparison of the Stokes\,$I$ emission, magnetic field segments, and polarized intensity $P$ of G31 at different wavelengths and angular resolution. ({\it Top panels}): Magnetic field segments ({\it red}) overlaid on the Stokes\,$I$ ({\it contours}) at ({\it left}) 879~$\mu$m observed with the SMA and a synthesized beam of 1$\farcs$34$\times$0$\farcs$83 by Girart et al.~(\cite{girart09}), ({\it middle}) 1.3\,mm with ALMA and a synthesized beam of 0$\farcs$28$\times$0$\farcs$20 by Beltr\'an et al.~(\cite{beltran19}), and ({\it right}) 1.3\,mmm with ALMA and a synthesized beam of 0$\farcs$072$\times$0$\farcs$064 (this work). Contours levels are $-$0.8\%, 0.8\%, 1.5\%, 2.5\%, 4\% and 6\% to 96\% in steps of 10\% of the peak intensity, which is 9.13\,Jy\,beam$^{-1}$ ({\it left}), $-$5, 5, 10, 15, 40, 160, and 300 times $\sigma$, where 1$\sigma$ is 1.2\,mJy\,beam$^{-1}$ ({\it middle}), and $-$5, 5, 10, 30, 60, 120, 160, and 200 times $\sigma$, where 1$\sigma$ is 0.15\,mJy\,beam$^{-1}$ ({\it right}). The synthesized beams are shown in the lower left corner. Black dots mark the position of four embedded continuum sources (\cite{beltran21}). ({\it Bottom panels}): Contour map of the Stokes\,$I$ emission superposed on the color image of the polarized flux intensity $P$.} 
\label{fig-history}
\end{center}
\end{figure*}

Another explanation could be that the polarized emission at 1.3\,mm is tracing the outer layers of the core, since the emission at this wavelength could be optically thicker, while the emission at 3.1\,mm is probing the deeper inner regions. To test this hypothesis, we have plotted the polarization fraction $p$ as a function of the normalized Stokes\,$I$, that is $I/I_{\rm max}$, where $I_{\rm max}$ is the maximum value of the intensity at each wavelength (Fig.~\ref{fig-I-polm}). As seen in this figure, the distributions at the two wavelengths are similar, with a slope consistent with $\sim$$-1$ in both cases.
Such a behavior has been observed in other regions (e.g., Lai et al.~\cite{lai03}; Tang et al.~\cite{tang09}) and could be produced by a loss of the efficiency of the polarization (Alves et al.~\cite{alves14}). Alternatively, it could also be caused by the presence of a more complex magnetic field toward the center of the core as a result of fragmentation and gravity, or by different filtering of the extended emission by the interferometer. In the latter case, Stokes\,$I$ would be more affected by filtering than Stokes\,$Q$ and $U$, and this would  artificially produce high values of the polarized fraction in regions where Stokes\,$I$ is weaker (Le Gouellec et al.~\cite{legouellec20}). 

\subsection{Mass-to-flux ratio}
\label{mass-to-flux}

We evaluated the mass-to-flux ratio, $\lambda$, using the expression
\be\label{m2fratio}
\lambda=2\pi G^{1/2}\frac{M(\Phi)}{\Phi}\,,
\ee
where $G$ is the gravitational constant, $\Phi$ the magnetic flux, and $M(\Phi)$ the mass contained in
the flux tube $\Phi$. 
We computed the magnetic flux, $\Phi=\pi R^2 B$, inside a radius of $1''$, corresponding to the Main core region, assuming spherical symmetry and a magnetic field strength of $\sim 13$\,mG, and obtained a  value of $1.3\times10^{32}$\,G\,cm$^2$. As for $M(\Phi)$, the mass contained in the flux tube $\Phi$, following Beltr\'an et al.~(\cite{beltran19}), we assumed that this is the mass of the core, 70\,$M_\sun$ (Cesaroni\,\cite{cesa19}), plus the mass of the (proto)stars already formed in the core, $\sim$20\,$M_\sun$ (Beltr\'an et al.~\cite{beltran19}). Consequently, $M(\Phi)=90$\,$M_\sun$ and this ``spherically averaged'' mass-to-flux ratio is $\lambda_s\sim 1.4$. To take into account that the mass of the core has been estimated inside a sphere of radius $\sim1\arcsec$ and not in a flux tube, as required by the definition of $\lambda$ in Eq.~(\ref{m2fratio}), we multiplied $\lambda_s$ by a correction factor equal to 1.4 (Li \& Shu~\cite{li96}) and obtained $\lambda \sim$1.9, which indicates that the core would be  ``supercritical''. 

On the other hand, applying the polarization-intensity gradient method described in Sect.~\ref{subsec:polint}, the ``spherically averaged'' mass-to-flux ratio $\lambda_s$ is given by
\begin{equation}
    \lambda_s = \frac{1}{\sqrt\pi} \left\langle \Sigma_B ^{-1/2} \right \rangle \left(\dfrac{R}{R_{\rm 0}}\right)^{-3/2}  ,
    \label{eq:lambdakoch}
\end{equation}
where $R_0 \equiv R$ is the cloud radius. Making use of the average value in map of $\left\langle \Sigma_B \right \rangle = 0.6$, we obtain in this case $\lambda_s = 1.45$, 
in agreement with the value of $\lambda_s$ derived above.

These values are consistent with the range $\lambda=1.4$--$2.2$ estimated by Beltr\'an et al.~(\cite{beltran19}), and with $\lambda=2.66$ ($\lambda_s=1.94$) assumed for the model. This mass-to-flux ratio should be taken as a lower limit because Beltr\'an et al.~(\cite{beltran19}) estimated the stellar mass content  assuming that the luminosity of 4.4$\times10^4$\,$L_\odot$ originates from a single star. However, the Main core is a massive protocluster that contains at least four embedded massive young stellar objects (Beltr\'an et al.~\cite{beltran21}), and therefore, the stellar content could be higher.  

We also estimated the mass-to-flux ratio in the inner part of the core surrounding sources A, B, and C, which corresponds to a radius  $R\sim0\farcs3$, and using a magnetic field strength of 78\,mG, which is the value obtained assuming a density of $5\times10^8$\,cm$^{-3}$ for the inner part of the Main core (see Sect.~\ref{subsec:fieldstrength}). To compute $M(\Phi)$, we added up the masses of sources A, B, and C,  57\,$M_\sun$, estimated by  Beltr\'an et al.~(\cite{beltran21}),  and added 20\,$M_\sun$ for the stellar content, for a total mass of 77\,$M_\sun$. In this case, $\Phi=0.7\times10^{32}$\,G\,cm$^{2}$, and $\lambda$, after correcting for the 1.4 factor, is 3.0, even more ``supercritical'', and suggesting that gravity dominates over magnetic field throughout the core.  

The Alfv\'en velocity was estimated following the expression
\be
\varv_{\rm A} = \frac{B}{\sqrt{4\pi\rho}}.
\ee
Using the estimates of the magnetic field strength in G31, we obtain an Alfv\'en velocity of $\sim$5.0\,\kms, for the two  values of $B$ (13--78\,mG) and $\rho$ ($1.4\times10^7$--$5\times10^8$\,cm$^{-3}$). Beltr\'an et al.~(\cite{beltran18}) estimated infall velocities of $\sim 2$--$8$\,\kms\ for the whole Main core from red-shifted absorption observed at $\sim0\farcs22$ resolution. The highest infall velocities are estimated for the vibrationally excited transitions of CH$_3$CN and for some transitions of the isotopologues $^{13}$CH$_3$CN and CH$_3^{13}$CN, which are optically thinner and should trace material close to the central (proto)star(s).  This has been confirmed by higher angular resolution observations ($\sim0\farcs09$) that have traced red-shifted absorption in H$_2$CO and CH$_3$CN (Beltr\'an et al.~\cite{beltran22}). The infall velocities estimated in H$_2$CO, which should trace more diffuse material in the outer part of the core, are $\sim$2--3\,\kms, while those estimated in CH$_3$CN, which should trace deeper embedded denser material, are $\sim$4--8\,\kms. This confirms the suggestion by Beltr\'an et al.~(\cite{beltran19}) that, while the collapse in the external part of the core is (slightly) sub-Alfv\'enic, it becomes super-Alfv\'enic close to the center.

The Davis-Chandrasekhar-Fermi method also allows to estimate the ratio of the turbulent component of the 
magnetic field, $\delta B\sim \sqrt{3}\,\delta B_{\rm los}$, to the uniform 
component $B=B_{\rm pos}/\cos i$. Using $B_{\rm pos}$ given by Eq.~(\ref{b-strength}) and $\sigma\sim \sqrt{3}\,\sigma_{\rm los}$, the $\delta B/B$ ratio results
\begin{equation}
\frac{\delta B}{B}\sim \frac{\sqrt{3}\cos i}{\xi}\delta\psi_{\rm int},
\end{equation}
for Alfv\'enic fluctuations, where $\delta$B$_{\rm los} = \sigma_{\rm los} \sqrt{4\pi\rho}$.

Using $\delta\psi_{\rm int}= 21.45^\circ$, $\delta B/B\sim0.8$, which indicates that the energy of the turbulent component of the
magnetic field in the Main core of G31 is a significant fraction ($\sim$60\% or larger) of the energy of the uniform component of the field included in our model.

\subsection{Magnetic field at different spatial scales}

The rotating and infalling HMC G31 is the first core in the high-mass regime for which polarized observations at $\sim$1$''$ ($\sim$3750\,au) and $\sim$0$\farcs$24 ($\sim$900\,au) angular resolutions, clearly reveal the characteristic hourglass shape of the magnetic field along its rotation axis (Girart et al.~\cite{girart09}; see Fig.~\ref{fig-history}).  As seen in Figs.~\ref{fig-angle-comparison} and \ref{fig-history}, the self-similarity observed at cores scales still holds at  circumstellar (disk/jet) scales, traced by observations at an angular resolution of  $\sim$0$\farcs$068 ($\sim$250\,au).
The modeling of the emission at core and circusmtellar scales confirms that the magnetic field is well represented by a poloidal field, with a negligible toroidal component despite the core shows a clear velocity gradient associated with rotation (e.g., Beltr\'an et al.~\cite{beltran18}). 

The magnetic field strength has been estimated at all scales with the DCF method, and with the polarization-intensity gradient method at the smallest scales,  and the values at core scales are all similar, with a value of $\sim$20\,mG at 1$''$, $\sim$8--13\,mG at 0$\farcs$24, and $\sim$13\,mG at 0$\farcs$068. Note that close to the center of the core, where the densities are $>10^8$\,cm$^{-3}$,  the magnetic field strength could be as high as $\sim$80\,mG. These values imply that the mass-to-flux ratio in this region is supercritical, with values ranging from $\sim$ 1.4 to 3.0 (Beltr\'an et al.~\cite{beltran19}; this work), and therefore, that gravity dominates over magnetic forces in the Main core. This is further supported by the detection of clear signatures of infall, such as red-shifted absorption, in the core (e.g., Girart et al.~\cite{girart09}; Mayen-Gijon et al.~\cite{mayen14}; Beltr\'an et al.~\cite{beltran18}; Estalella et al.~\cite{estalella19}). 

The velocity measurements at the largest scales indicate that 
the collapse of the core is sub-Alfv\'enic (Girart et al.~\cite{girart09}). The 0$\farcs$2 observations confirm that the collapse in the external part of the core is slightly  sub-Alfv\'enic but it becomes super-Alfv\'enic toward the center of the core. This behaviour is confirmed by the observations at the highest angular resolution (see Sect.~\ref{mass-to-flux}). 

Finally, observations suggest that while at large scales magnetic energy dominates over turbulent energy (Girart et al.~\cite{girart09}), at smaller scales the turbulent component of the magnetic field in the Main core of G31 is a significant fraction (60\% or larger) of the energy of the uniform component.

All these findings point to a self-similarity of the magnetic field from large (core) to small (circumstellar) scales.  However, the fact that the dust emission at 1.3\,mm might be optically thick, as indicated by the detection of dichroic  extinction in the inner region of the core, suggests that despite the high-angular resolution ($\sim$0$\farcs$068 or $\sim$250\,au) of our observations, we might not be properly tracing the inner region of the core, close to the embedded protostars. For this reason our observations, despite their high angular resolution, might be hindered by the large opacity of the central region of the core and thus be
unsuited to unveil the magnetic field configuration on circumstellar scales.
This could explain why, even on scales of a few 100s au, the magnetic field appears to have no toroidal component, as one would expect at these scales. Unfortunately, we cannot study the magnetic field morphology toward the (proto)stars even at 3.1\,mm because, despite the emission at this wavelength being optically thinner, the observations are less sensitive than at 1.3\,mm and the core at 3.1\,mm is not properly sampled. Hence the data are insufficient  
to properly constrain the models, which are axisymmetric (see Fig.~\ref{fig-iphichi2-3mm}), because inclination and position angles can be either positive or negative. Besides the sensitivity problems and the non-uniform sampling of the field, the 3.1\,mm high-angular resolution observations by peering inside the core could be tracing a more perturbed field, as a result of the presence of molecular outflows associated with all the embedded sources, in which polarization efficiency and grain alignment might not be as good as at low angular resolution.  

In conclusion, very high angular resolution polarized observations at short wavelengths ($\lesssim3$\,mm) might be unsuited to study the magnetic field in dense cores, due to the fact that the emission might be too optically thick. A solution could be to perform sensitive observations at longer wavelengths, such as those offered by the new ALMA Band\,1 receiver. Only by adequately resolving the dust emission, we will be able to investigate the magnetic field properties around each of the collapsing and embedded protostars in G31 and establish the true configuration of it at circumstellar scales.

\section{Conclusions}

We carried out ALMA 1.3\,mm and 3.1\,mm polarization observations at $\sim0\farcs068$ ($\sim$250\,au) angular resolution of the HMC  G31.41+0.31, previously observed with the SMA at 870\,$\mu$m with 1$''$ resolution and ALMA at 1.3\,mm with 0$\farcs$24 resolution. The aims of this study were to investigate the morphology of the magnetic field at both wavelengths, estimate the magnetic field strength using different methods, the DCF and the polarization-intensity gradient method, model the magnetic field and examine whether the self-similarity observed at core scales still holds at circumstellar (disk/jet) scales.

While the distributions of the polarized emission at the two wavelengths is basically the same,  the polarized intensity is weaker at 3.1\,mm than at 1.3\,mm. The polarized emission is associated with the four sources, A, B, C, and D, embedded in the core. The orientation of the magnetic field obtained at 1.3\,mm coincides with that and 3.1\,mm, except for an area between sources C and B.  Such an agreement indicates that there is no wavelength dependence and that the polarized observations are probably tracing the emission of magnetically aligned grains and are not affected by dust self-scattering. The optically thick 1.3\,mm emission observed between sources B and C could be affected by dichroic extinction confined to a small area inside G31, suggesting that very high optical depths are needed to favour this mechanism over magnetic alignment of the dust grains.

The polarized emission  has been successfully modeled with the same semi-analytical magnetostatic model of a self-gravitating toroid supported by magnetic fields used in Beltr\'an et al.~(\cite{beltran19}). The best fit model suggests that the magnetic field associated with G31 is well represented by a poloidal field with a small toroidal component on the order of 10\% of the poloidal component ($b_0 = 0.1$),  with an axis oriented SE--NW at position angle $\phi=-63^\circ$ and inclination $i = 50^\circ$, for a mass-to-flux ratio $\lambda = 2.66$. As already found from previous lower angular resolution observations, the magnetic field axis is oriented perpendicular to the NE--SW velocity gradient detected in this core on scales from $\sim10^3$ to 10$^4$\,au.

The magnetic field strength in the plane-of-the-sky has been estimated using two different methods, the DCF and the polarization-intensity gradient methods. Both indicate values in the range $\sim$10--$80$\,mG for a density range $1.4 \times 10^7 - 5 \times 10^8$\,cm$^{-3}$. These values of the magnetic field are consistent with those expected from the distribution of magnetic field strength as a function of density of Crutcher~(\cite{crutcher12}). 

The mass-to-flux ratio estimated is in the range $\lambda\sim1.9$--3.0, using both the DCF and the polarization-intensity gradient methods, which indicates that the core is ``supercritical''. This is consistent with the analysis of the relative importance of magnetic and gravitational forces obtained with the polarization-intensity gradient method, which indicates that the magnetic field is too weak to prevent gravitational collapse inside the G31 core. The collapse in the external part of the core is (slightly) sub-Alfvénic but becomes super-Alfvénic close to the center.

The new ALMA high-angular resolution observations at 1.3\,mm and 3.1\,mm have confirmed the hourglass-shaped magnetic field morphology observed previously with the SMA at 1$''$ and with ALMA at $0\farcs24$.  This suggests a self-similarity of the magnetic field from large (core) to small (circumstellar) scales. However, we cannot discard the possibility that the observations at both 1.3\,mm and 3.1\,mm are partially optically thick and as such could not be suitable to properly trace the emission, and thus, the magnetic field, at the smallest scales. Sensitive high-angular resolution observations at longer wavelengths, such as those offered by the new ALMA Band\,1 receiver, should allow us to properly resolve the emission at circumstellar scales and investigate the magnetic field properties around each of the accreting embedded protostars in G31.
 
\begin{acknowledgements}

M.T.B., M.P., D.G., R.C., D.D.O., and L.M.\ acknowledge financial support through the INAF Large Grant {\it The role of MAGnetic fields in MAssive star formation} (MAGMA).  N.A.L.\ acknowledges financial support from the European Research Council (ERC) via the ERC Synergy Grant ECOGAL (grant 855130). J.M.G.\ acknowledges support by the grant PID2020-117710GB-I00 (MCI-AEI-FEDER, UE).  A.S.-M.\ acknowledges support from the RyC2021-032892-I grant funded by MCIN/AEI/10.13039/501100011033 and by the European Union `Next GenerationEU’/PRTR, and support from the PID2020-117710GB-I00 (MCI-AEI-FEDER, UE). J.M.G.\ and A.S.-M.\  are also partially supported by the program Unidad de Excelencia Mar\'{\i}a de Maeztu CEX2020-001058-M. G.A.\ and M.O.\ acknowledge financial support from grants PID2020-114461GB-I00 and CEX2021-001131-S, funded by MCIN/AEI/10.13039/501100011033.  This paper makes use of the following ALMA data: ADS/JAO.ALMA\#2018.1.00632.S. ALMA is a partnership of ESO (representing its member states), NSF (USA) and NINS (Japan), together with NRC (Canada), MOST and ASIAA (Taiwan), and KASI (Republic of Korea), in cooperation with the Republic of Chile. The Joint ALMA Observatory is operated by ESO, AUI/NRAO and NAOJ.

\end{acknowledgements}

\appendix

\section{Dichroic extinction}

\begin{figure}
\begin{center}
\includegraphics[angle=0,width=8.5cm]{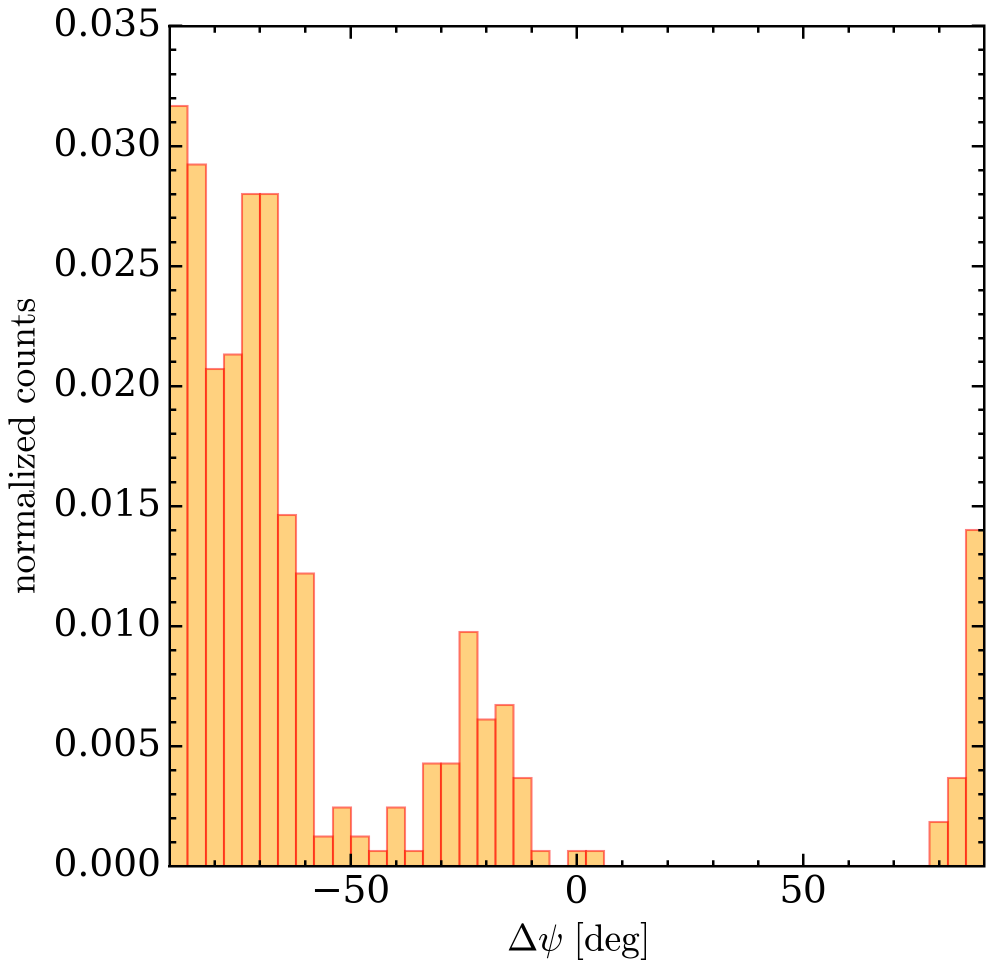} 
\vspace{-2.5cm}
\caption{
Histogram of the polarization angle difference between the 1.3\,mm and 3.1\,mm observations in the area between sources B and C.} 
\label{fig-histo-dichroic}
\end{center}
\end{figure}

Figure~\ref{fig-histo-dichroic} shows the histogram of the average difference between the 
polarization angles at 1.3\,mm and 3.1\,mm only for the area between sources B and C, which could be affected by dichroic extinction. As seen in this figure, the average difference between the polarization angles at
the two wavelengths $\Delta\psi$ is $\gtrsim75^\circ$ or $\lesssim -75^\circ$.

\section{{\tt uv}-tapering}

\begin{figure}
\begin{center}
\includegraphics[angle=0,width=8.5cm]{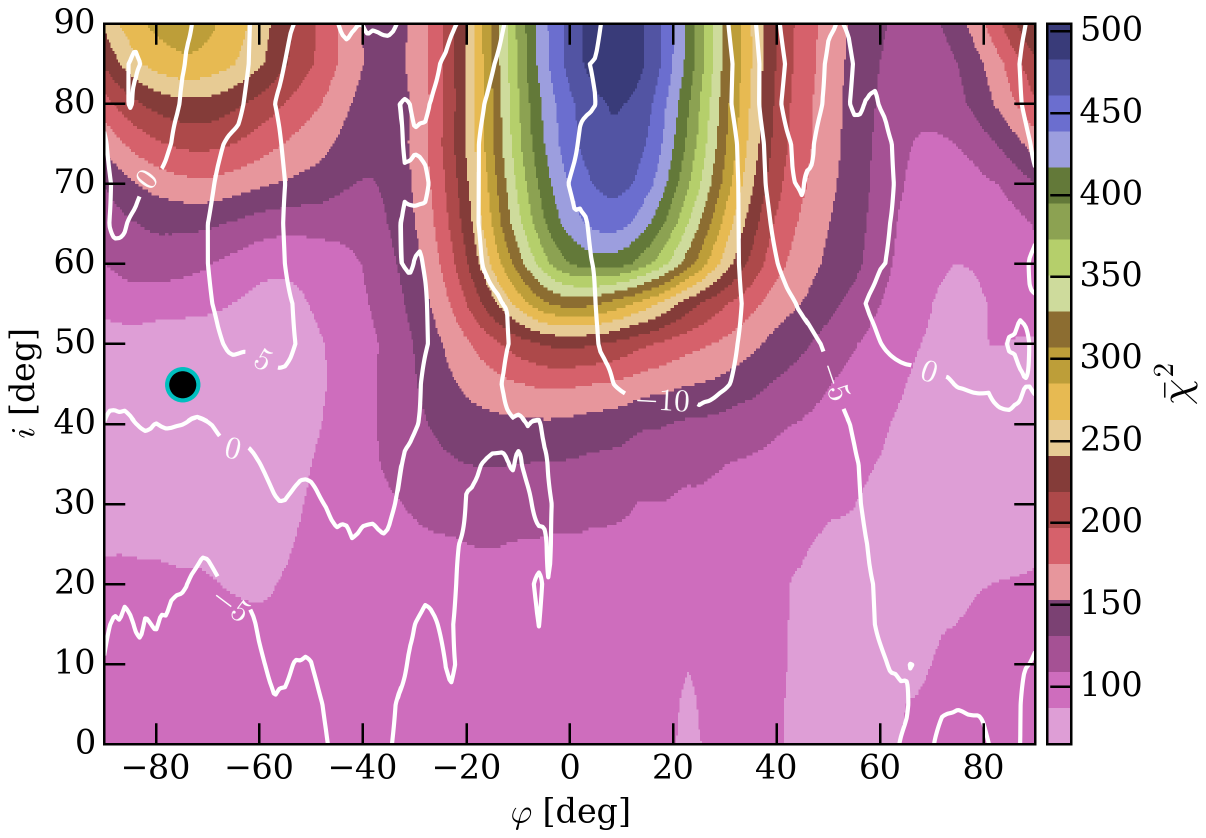}
\vspace{-3cm}
\caption{Same as Fig.~\ref{fig-iphichi2-1mm}, but for 3.1 mm observations with {\tt uv}-tapering.}
\label{fig-iphichi2-3mmuvtaper}
\end{center}
\end{figure}

To possibly increase the signal-to-noise of the 3.1\,mm observations and check whether we could recover some low-intensity extended emission, we applied a {\tt uv}-taper of $0\farcs15\times0\farcs15$ to the visibility data when running {\tt tclean}. The resulting synthesized beam of the maps is $0\farcs19\times0\farcs16$. Figure~\ref{fig-iphichi2-3mmuvtaper} shows the distribution of the reduced chi-squared values $\bar\chi^2$ between the model and the observations, after performing the {\tt uv}-tapering, as a function of the inclination $i$ and the orientation of the projection of the magnetic axis on the plane of the sky $\varphi$.

\begin{thebibliography}{}

\bibitem[2014]{alves14}
Alves, F.\ O., Frau, P., Girart, J.\ M., et al.\ 2014, A\&A, 569, L1 

\bibitem[2018]{alves18}
Alves, F.\ O., Girart, J.\ M., Padovani, M., Galli, D.\ et al.\ 2018, A\&A, 616, A56

\bibitem[2018]{bacciotti18}
Bacciotti, F., Girart, J.\ M., Padovani, M., Podio, L.\ et al.\ 2018, ApJ, 865, L12


\bibitem[2004]{beltran04} 
Beltr\'an, M.\ T., Cesaroni, R., Neri, R., Codella, C., Furuya, R.\ S., Testi, L., \& Olmi, O.\ 2004, ApJ, 601, L190 


\bibitem[2018]{beltran18} 
Beltr\'an, M.\ T., Cesaroni, R., Rivilla, V.\ M., S\'anchez-Monge, \'A.\ et al.\  2018, A\&A, 615, A141  

\bibitem[2016]{beltran16} 
Beltr\'an, M.\ T., \& de Wit, W.\ J.\ 2016, A\&ARv, 24, 6

\bibitem[2019]{beltran19} 
Beltr\'an, M.\ T., Padovani, M., Girart, J.\ M., Galli, D.\ et al.\  2019, A\&A, 630, A54  

\bibitem[2022]{beltran22} 
Beltr\'an, M.\ T., Rivilla, V.\ M., Cesaroni, R.,  Galli, D.\ et al.\ 2022, A\&A, 659, A81  

\bibitem[2021]{beltran21} 
Beltr\'an, M.\ T., Rivilla, V.\ M., Cesaroni, R.,  Maud, L.\ T.\ et al.\ 2021, A\&A, 648, A100  

\bibitem[1995]{briggs95}
Briggs, D.\ 1995, PhD Thesis, New Mexico Inst., Mining \& Tech.


\bibitem[2019]{cesa19}
Cesaroni, R.\ 2019, A\&A, 631, A65

\bibitem[1994]{cesa94}
Cesaroni, R., Churchwell, E., Hofner, P., Walmsley, C.\ M., \& Kurtz, S.\ 1994, A\&A, 288, 903

\bibitem[2010]{cesa10}
Cesaroni, R., Hofner, P., Araya, E., \& Kurtz, S.\ 2010, A\&A, 590, A50

\bibitem[1953]{chandra53}
Chandrasekhar, S. \& Fermi, E. 1953 ApJ, 118, 113

\bibitem[2022]{commercon22}
Commer\c{c}on, B., Gonz\'alez, M., Mignon-Risse, R., et al.\ 2022, A\&A, 658, A52

\bibitem[2011]{commercon11}
Commer\c{c}on, B., Hennebelle, P., \& Henning, T. 2011, ApJ, 742, L9 

\bibitem[2012]{crutcher12}
Cruchter, R.\ M.\ 2012, ARA\&A, 50, 29 


\bibitem[2001]{dalessio01}
D'Alessio, P.,  Calvet, N., \& Hartmann, L.\ 2001, ApJ, 553, 321

\bibitem[2014]{davidson14}
Davidson, J.~A., Li, Z.-Y., Hull, C.~L.~H., et al.\ 2014, \apj, 797, 74 

\bibitem[1951]{davis51}
Davis, L.\ 1951, Phys.\ Rev.\ 81, 890

\bibitem[2018]{dent18}
Dent, W.\ R.\ F., Pinte, C., Cortes, P.\ C., M\'enard, F.\ et al.\ 2018, \mnras, 482, L29

\bibitem[2019]{estalella19}
Estalella, R., Anglada, G., D\'{\i}az-Rodr\'{\i}guez, A.\ K., Mayen-Gijon, J.\ M.\ 2019, A\&A, 626, 84 

\bibitem[1993]{fiedler93}
Fiedler, R.\ A., \& Mouschovias, T.\ Ch.\ 1993, ApJ, 415, 680

\bibitem[2011]{frau11}
Frau, P., Galli, D., \& Girart, J.\ M.\ 2011, A\&A, 535, 44


\bibitem[1993a]{galli93a}
Galli, D., \& Shu, F.\ H.\ 1993a, ApJ, 417, 220

\bibitem[1993b]{galli93b}
Galli, D., \& Shu, F.\ H. \ 1993b, ApJ, 417, 243


\bibitem[1999]{girart99}
Girart, J.~M., Crutcher, R.~M., \& Rao, R.\ 1999, \apjl, 525, L109 

\bibitem[2009]{girart09}
Girart, J.\ M., Beltr\'an, M.\ T., Zhang, Q., Rao, R., \& Estalella, R.\ 2009, Science, 324, 1408 

\bibitem[2018]{girart18}
Girart, J.\ M., Fern\'andez-L\'opez, M., Li, Z.-Y., Yang, H.\ et al.\ 2018, ApJ, 856, L27

\bibitem[2006]{girart06}
Girart, J.\ M., Rao, R., Marrone, D.\ P.\ 2006, Science, 313, 812

\bibitem[2008]{goncalves08}
Gon\c{c}alves, J., Galli, D., \& Girart, J.\ M.\ 2008, A\&A, 490, L39

\bibitem[2000]{hildebrand00}
Hildebrand, R.\ H., Davidson, J.\ A., Dotson, J.\ L., et al.\ 2000, PASP, 112, 1215





\bibitem[2020]{hull20}
Hull, C.~L.~H., Cortes, P.\ C., Le Gouellec, V.\ J.\ M.\ et al., 2020, PASP, 132, 094501

\bibitem[2014]{hull14}
Hull, C.~L.~H., Plambeck, R.\ L., Kwon, W., et al.\ 2014, ApJs, 213, 13

\bibitem[2018]{hull18}
Hull, C.~L.~H., Yang, H., Li, Z.-Y., Kataoka, A.\ et al.\ 2018, ApJ, 860, 82

\bibitem[2019]{hull19}
Hull, C.~L.~H., \& Zhang, Q.\ 2019, Frontiers in Astronomy and Space Sciences, 6, 3

\bibitem[2019]{immer19}
Immer, K., Li, J., Quiroga-Nu\~nez, L.\ H., et al.\ 2019, A\&A, 632, A123

\bibitem[2015]{kataoka15}
Kataoka A., Muto, T., Momose, M., Tsukagoshi, T.\ et al.\ 2015, ApJ, 809, 78

\bibitem[2017]{kataoka17} 
Kataoka, A., Tsukagoshi, T., Pohl, A., et al.\ 2017, \apj, 844, L5

\bibitem[2020]{ko20}
Ko, C.-L, Liu, H.\ B., Lai, S.-P., et al.\ 2020, ApJ, 889, 172

\bibitem[2012a]{koch12a}
Koch, P.~M., Tang, Y.-W., \& Ho, P.~T.~P., et al.\ 2012a, \apj, 747, 79

\bibitem[2012b]{koch12b}
Koch, P.~M., Tang, Y.-W., \& Ho, P.~T.~P., et al.\ 2012b, \apj, 747, 80

\bibitem[2013]{koch13}
Koch, P.~M., Tang, Y.-W., \& Ho, P.~T.~P., et al.\ 2013, \apj, 775, 77

\bibitem[2014]{koch14}
Koch, P.~M., Tang, Y.-W., Ho, P.~T.~P., et al.\ 2014, \apj, 797, 99 

\bibitem[2018]{kwon18}
Kwon, W., Stephens, I., Tobin, J., et al.\ 2018, arXiv:1805.07348 

\bibitem[2003]{lai03}
Lai, Sh.\ P., Girart, J.\ M., \& Crutcher, R.\ M.\ 2003, ApJ, 598, 392 

\bibitem[2015]{lazarian15}
Lazarian, A., Andersson, B.\ G.,, \& Hoang, T.\ 2015, in Polarimetry of stars and planetary systems, 81


\bibitem[2020]{legouellec20}
Le Gouellec, V.\ J.\ M., Maury, A.\ J., Guillet, V., et al.\ 2020, A\&A, 644, A11 

\bibitem[1996]{li96}
Li, Z.-Y., \& Shu, F.~H.\ 1996, ApJ, 472, 211


\bibitem[2015]{li15}
Li, H.-B., Yuen, K.\ H., Otto, F., Leung, P.\ K.\ et al.\ 2015, Nature, 520, 518

\bibitem[2023]{lin23}
Lin, Z.-Y.~D., Li, Z.-Y., Stephens, I.~W., et al.\ 2023, arXiv:2309.10055. doi:10.48550/arXiv.2309.10055

\bibitem[2021]{liu21}
Liu, H.\ B.\ 2021, ApJ, 914, 25

\bibitem[2021]{liu-etal21}
Liu, J.,  Zhang, Q., Commer\c {c}on, B., et al.\ 2021, ApJ, 919, 79

\bibitem[2022]{liu22}
Liu, J.,  Qiu, K., \& Zhang, Q.\ 2022, ApJ, 925, 30



\bibitem[1994]{mannings94}
Mannings, V., \& Emerson, J.\ P.\ 1994, MNRAS, 267, 361 

\bibitem[2014]{mayen14}
Mayen-Gijon, J.\ M., Anglada, G., Osorio, M., Rodr\'{\i}guez, L.\ F., Lizano, S.\ et al.\  2014, MNRAS, 437, 3766

\bibitem[2018]{maury18}
Maury, A.~J., Girart, J.~M., Zhang, Q., et al.\ 2018, \mnras, 477, 2760 

\bibitem[2007]{mcmullin07}
McMullin, J.\ P., Waters, B., Schiebel, D., Young, W., \& Golap, K.\ 2007, ASPC, 376,  127	

\bibitem[2006]{mouschovias06}
Mouschovias, T.\ Ch., Tassis, K., \& Kunz, M.\ W.\ 2006, ApJ, 646, 1043

\bibitem[2009]{osorio09}
Osorio, M., Anglada, G., Lizano, S., \& D'Alessio, P.\ 2009, ApJ, 694, 29

\bibitem[1994]{ossenkopf94}
Ossenkopf, V., \& Henning, Th.\ 1994, A\&A, 291, 943

\bibitem[2001]{ostriker01}
Ostriker, E.~C., Stone, J.~M., \& Gammie, C.~F.
2001, ApJ, 546, 980

\bibitem[2011]{padovani11}
Padovani, M. \& Galli, D., 2011, A\&A, 530, A109

\bibitem[2012]{padovani12}
Padovani, M., Brinch, C., Girart, J.~M., J\o rgensen, J.~K., Frau, P. et al.\ 2012, A\&A, 543, A16



\bibitem[2013]{padovani13}

\bibitem[1994]{pollack94}
Pollack, J.\ B., Hollenbach, D., Beckwith, S., et al.\ 1994, ApJ, 421, 615

\bibitem[2014]{qiu14} 
Qiu, K., Zhang, Q., Menten, K.~M., et al.\ 2014, \apjl, 794, L18 

\bibitem[1998]{rao98}
Rao, R., Crutcher, R.\ M., Plambeck, R.\ L., \& Wright, M.\ C.\ H.\ 1998, ApJ, 502, L75 



\bibitem[2013]{sadavoy13}
Sadavoy, S. I., Di Francesco, J., Johnstone, D., et al.\ 2013, ApJ, 767, 126

\bibitem[1998]{schleuning98} 
Schleuning, D.~A.\ 1998, \apj, 493, 811 


\bibitem[2014]{tan14}
Tan, J.\ C., Beltr\'an, M.\ T., Caselli, P., et al.\ 2014, in Protostars and Planets VI, eds. H.\ Beuther, R.\ S.\ Klessen, C.\ P.\ Dullemond, \& Th.\ Henning, University of Arizona Press, Tucson, 149

\bibitem[2009]{tang09}
Tang, Y.-W., Ho, P.\ T.\ P.,Koch, P.\ M., et al.\ 2009, ApJ, 700, 251

\bibitem[2019]{tang19}
Tang, Y.-W., Koch, P.\ M., Peretto, N., et al.\ 2019, ApJ, 878, 10

\bibitem[2006]{vaillancourt06}
Vaillancourt, J.\ E.\ 2006, PASP 118, 1340

\bibitem[1997]{wood97}
Wood, K.\ 1997, ApJ, 477, L25

\bibitem[2017]{yang17}
Yang, H., Li, Z.-Y, Looney, L.\ W., Girart, J.\ M., \& Stephens, I.\ W.\ 2017, MNRAS, 472, 373

\bibitem[2014]{zhang14} 
Zhang, Q., Qiu, K., Girart, J.~M., et al.\ 2014, \apj, 792, 116 


\end{thebibliography}
\end{document}